\documentclass[journal=ancac3,manuscript=article]{achemso} 

\usepackage[version=3]{mhchem} 
\usepackage{graphicx}
\usepackage{amsmath, amssymb, bm}
\usepackage{booktabs}
\usepackage{siunitx}
\usepackage{dcolumn}
\usepackage{xcolor}
\usepackage{wrapfig} 
\usepackage{tabularx}
\usepackage{multirow}
\usepackage{nicefrac}
\usepackage{ragged2e} 
\usepackage{subcaption}

\usepackage[colorlinks=true,citecolor=blue,linkcolor=blue]{hyperref}

\usepackage[normalem]{ulem}
\newcommand{\editor}[2]{%
  \expandafter\newcommand\csname #1note\endcsname[1]{\textcolor{#2}{(\textbf{#1:} ##1)}}%
  \expandafter\newcommand\csname #1\endcsname[1]{\textcolor{#2}{##1}}%
  \expandafter\newcommand\csname #1cancel\endcsname[1]{\textcolor{#2}{\sout{##1}}}%
  \expandafter\newcommand\csname #1change\endcsname[2]{\textcolor{#2}{\sout{##1} ##2}}%
  \newenvironment{#1text}{\color{#2}}{\color{black}}
}

\usepackage{cleveref}

\editor{FH}{blue}
\editor{MG}{orange}
\editor{R}{red}

\setcitestyle{super}
\title{Exploring the magnetic landscape of easily-exfoliable two-dimensional materials}

\author{Fatemeh Haddadi} 
\affiliation{Theory and Simulation of Materials (THEOS), \'Ecole Polytechnique F\'ed\'erale de Lausanne (EPFL), CH-1015 Lausanne, Switzerland}
\alsoaffiliation{National Centre for Computational Design and Discovery of Novel Materials (MARVEL), \'Ecole Polytechnique F\'ed\'erale de Lausanne (EPFL), CH-1015 Lausanne, Switzerland}
\email{fatemeh.haddadi@epfl.ch}

\author{Davide Campi}
\affiliation{Department of Materials Science, University of Milano-Bicocca, Via R. Cozzi 55, Milano 20125, Italy}

\author{Flaviano dos Santos}
\affiliation{Laboratory for Materials Simulations (LMS), Paul Scherrer Institut, Villigen PSI, Switzerland}
\alsoaffiliation{National Centre for Computational Design and Discovery of Novel Materials (MARVEL), \'Ecole Polytechnique F\'ed\'erale de Lausanne (EPFL), CH-1015 Lausanne, Switzerland}

\author{Nicolas Mounet}
\affiliation{Theory and Simulation of Materials (THEOS), \'Ecole Polytechnique F\'ed\'erale de Lausanne (EPFL), CH-1015 Lausanne, Switzerland}
\alsoaffiliation{National Centre for Computational Design and Discovery of Novel Materials (MARVEL), \'Ecole Polytechnique F\'ed\'erale de Lausanne (EPFL), CH-1015 Lausanne, Switzerland}
\altaffiliation{Now at: CERN (European Organization for Nuclear Research), Geneva, Switzerland}

\author{Louis Ponet}
\affiliation{Theory and Simulation of Materials (THEOS), \'Ecole Polytechnique F\'ed\'erale de Lausanne (EPFL), CH-1015 Lausanne, Switzerland}
\alsoaffiliation{National Centre for Computational Design and Discovery of Novel Materials (MARVEL), \'Ecole Polytechnique F\'ed\'erale de Lausanne (EPFL), CH-1015 Lausanne, Switzerland}

\author{Nicola Marzari}
\affiliation{Theory and Simulation of Materials (THEOS), \'Ecole Polytechnique F\'ed\'erale de Lausanne (EPFL), CH-1015 Lausanne, Switzerland}
\alsoaffiliation{National Centre for Computational Design and Discovery of Novel Materials (MARVEL), \'Ecole Polytechnique F\'ed\'erale de Lausanne (EPFL), CH-1015 Lausanne, Switzerland}

\author{Marco Gibertini}
\affiliation{Dipartimento di Scienze Fisiche, Informatiche e Matematiche, University of Modena and Reggio Emilia, I-41125 Modena, Italy}
\alsoaffiliation{Centro S3, CNR-Istituto Nanoscienze, I-41125 Modena, Italy}
\email{marco.gibertini@unimore.it}

\keywords{two-dimensional magnetic materials, hubbard corrections, global and local minima, energy landscape, density-functional theory, DFT+$U$}

\begin{document}
\begin{abstract}
Magnetic materials often exhibit complex energy landscapes with multiple local minima, each corresponding to a self-consistent electronic structure solution. Finding the global minimum is challenging, and heuristic methods are not always guaranteed to succeed. Here, we apply a recently developed automated workflow to systematically explore the energy landscape of 194 magnetic monolayers obtained from the Materials Cloud 2D crystals database and determine their ground-state magnetic order. Our approach enables effective control and sampling of orbital occupation matrices, allowing rapid identification of local minima. We find a diverse set of self-consistent collinear metastable states, further enriched by Hubbard-corrected energy functionals, when the $U$ parameters have been computed from first principles using linear-response theory. We categorise the monolayers by their magnetic ordering and highlight promising candidates. Our results include 109 ferromagnetic, 83 antiferromagnetic, and 2 altermagnetic monolayers, along with 12 novel ferromagnetic half-metals with potential for spintronics technologies.
\end{abstract}
\crefname{figure}{Figure}{Figures}


\maketitle


\label{sec1}
Magnetism in two-dimensional (2D) van der Waals (vdW) materials has garnered significant attention in recent years,~\cite{burch_magnetism_2018, wang_magnetic_2022,rhone_artificial_2023,gibertini_magnetic_2019,sodequist_type_2023,ovesen_orbital_2024,xin_machine_2023,shen_high-throughput_2022,xia_accelerating_2022} not only for potential technological applications but also for the large variety of phenomena that can be hosted by magnetic 2D materials. Indeed, magnetism in 2D displays very different properties depending on the interplay between thermal fluctuations, exchange interactions, and magnetic anisotropy. The thrust to investigate this richness in experiments has stimulated an intense search for novel magnetic monolayers displaying specific phenomena, from Ising anisotropy,~\cite{huang_layer-dependent_2017} to Berezinskii-Kosterlitz-Thouless transitions,~\cite{bedoya-pinto_intrinsic_2021} and heavy-fermion excitations~\cite{posey_two-dimensional_2024}. 

Although many candidates have been suggested by a long-term experience with layered magnetic materials, the continuous demand for novel 2D magnets calls for more systematic and exhaustive investigations. In this respect, theoretical approaches, particularly first-principles calculations based on density functional theory (DFT), provide a robust and high-throughput approach to predict novel 2D materials,~\cite{mounet_two-dimensional_2018,campi_expansion_2023, lebegue_two-dimensional_2013,rasmussen_computational_2015, choudhary_high-throughput_2017, ashton_topology-scaling_2017, cheon_data_2017, zhou_2dmatpedia_2019, haastrup_computational_2018, gjerding_recent_2021} including magnetic monolayers~\cite{mounet_two-dimensional_2018,torelli_high_2019,kabiraj_high-throughput_2020} and their properties.~\cite{torelli_first_2020, torelli_calculating_2018} 

Despite many successes, DFT can sometimes struggle to accurately predict the magnetic properties of materials, owing to the localized nature of $d$- and $f$-orbitals typically involved in magnetism and their tendency to suffer from self-interaction errors when adopting approximate exchange-correlation functionals. A computationally inexpensive and effective strategy to overcome these limitations is achieved by complementing approximate functionals with self-interaction corrections~\cite{kulik_density_2006} in the so-called Hubbard DFT+$U$ approach.~\cite{anisimov_band_1991, anisimov_first-principles_1997, dudarev_electron-energy-loss_1998} The Hubbard $U$ parameter entering the DFT+$U$ approach can be considered, somehow unsatisfactory, as a semi-empirical parameter, or can be computed self-consistently, e.g.\ using linear-response theory~\cite{cococcioni_linear_2005}. In the latter case, the theory preserves a first-principles character, and efficient implementations using density-functional perturbations theory (DFPT) are suitable for high-throughput investigations.~\cite{moore_high-throughput_2022,timrov_self-consistent_2021, timrov_hp_2022, timrov_hubbard_2018,cococcioni_linear_2005,bastonero_first-principles_2025,dudarev_electron-energy-loss_1998,leiria_campo_jr_extended_2010, haddadi_-site_2024} 

Another crucial challenge for simulations is the identification of the ground state magnetic configuration of the system. Indeed, the magnetic energy landscape is typically complex, with several local minima associated with different magnetic states. The inclusion of Hubbard corrections, needed for a better description of magnetic systems, typically leads to a proliferation of local minima, making it even more challenging to identify the true ground state (global minimum).\cite{meredig_method_2010,zhou_obtaining_2009, ylvisaker_anisotropy_2009,jollet_hybrid_2009,jomard_structural_2008,amadon__2008,zhang_magnetic_2009,kasinathan_ferromagnetism_2007,shick_spin_2004,ponet_energy_2024} Several approaches have been developed to tackle this issue, ranging from efficiently generating different magnetic configurations,~\cite{horton_high-throughput_2019} to the manipulation of the occupation matrix for $d$- or $f$-orbitals,~\cite{payne_optimizing_2019,allen_occupation_2014,dorado_stability_2010,dorado_dft_2009,meredig_method_2010} to computing spin waves~\cite{tellez-mora_systematic_2024} and spin spirals~\cite{sodequist_magnetic_2024},  employing cluster multiple theory,~\cite{huebsch_benchmark_2021} genetic algorithms~\cite{zheng_maggene_2021} or machine learning methods.\cite{baumsteiger_exploring_2025}

In this work, we thus explore the magnetic properties of easily exfoliable 2D materials from the Materials Cloud two-dimensional crystals database (MC2D)~\cite{campi_expansion_2023, mounet_two-dimensional_2018, campi_materials_2022, mc2d} using our recently developed approaches to reveal novel magnetic monolayers. Starting from 3077 easily and potentially exfoliabble monolayers (2004 easily and 1073 potentially exfoliable) with up to 40 atoms per cell~\cite{campi_expansion_2023, mounet_two-dimensional_2018}, 877 systems with up to 12 atoms per cell are screened with plain DFT using an AiiDA\cite{pizzi_aiida_2016,huber_aiida_2020} workflow (Chronos)\cite{mounet_two-dimensional_2018}. Out of these systems, 483 monolayers are found to be non-magnetic, 166 systems were discarded because of the failure in the optimization, and 228 monolayers showed a magnetic ground state. For these 228 materials, Hubbard corrections are then included for better accuracy, with the Hubbard parameter computed self-consistently. To identify the magnetic ground state, we employ the Robust Occupation Matrix Energy Optimization (RomeoDFT) algorithm~\cite{ponet_energy_2024} based on constraining the atomic orbital occupation matrices that allows us to systematically explore the magnetic energy landscape. With this protocol, we identify 194 magnetic 2D materials for which we accurately determine the magnetic ground state and investigate electronic and magnetic properties, including a proxy for the critical temperature. Notably, we identify 12 half-metals, a class of materials with spin-dependent conductivity, where electrons in one spin channel (up or down) are metallic, while those in the opposite channel are insulating, which is especially promising for spintronics applications. The remaining 34 systems were discarded because either they were repeated, found non-magnetic, consistently failed in the calculation of the Hubbard parameter, or not enough minima were found using RomeoDFT (see Supporting Information). \\

\begin{figure*}
    \centering\includegraphics[width=1\textwidth]{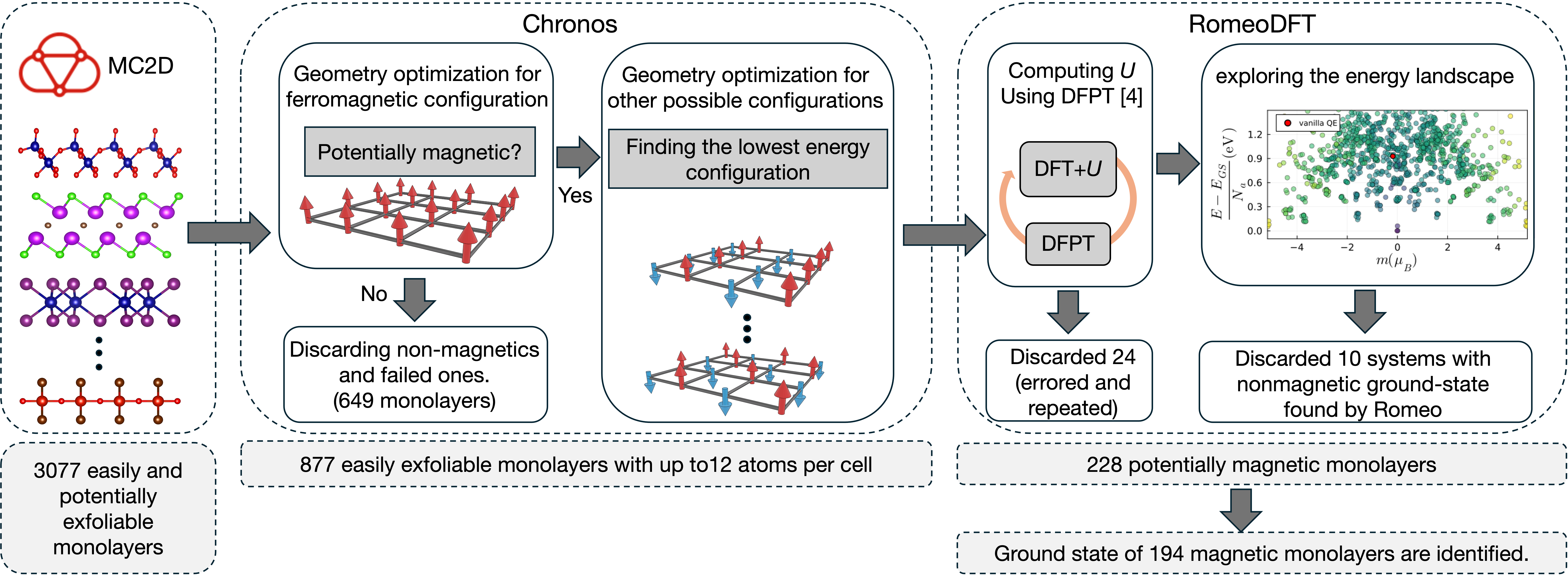}
    \caption{\label{workflow}
    The workflow for identifying the ground state of magnetic monolayers in this work that is done in two main steps. In the first step, 877 easily exfoliable monolayers from MC2D~\cite{campi_expansion_2023, mounet_two-dimensional_2018, campi_materials_2022, mc2d} are screened. Materials are identified as potentially magnetic if the total energy of the optimized geometry in the ferromagnetic ordering is smaller than the non-magnetic state. Then, more magnetic configurations, including antiferromagnetic ordering, are explored, making sure there are two magnetic atoms in the unit cell to accommodate antiferromagnetic ordering and the geometry is optimized starting from different starting magnetization (defined by the spin density). 228 magnetic monolayers with up to 12 atoms per unit cell are identified and go to the next step. The Hubbard $U$ is computed self-consistently in three consecutive steps using linear-response theory within DFPT. 24 systems that are repeated or failed in the computation of Hubbard parameters are discarded. After that, the geometry and Hubbard $U$ are fixed, and the energy landscape is explored by constraining the occupation matrix of $d$ or $f$ orbitals and calculating the total energy of different minima of the energy curvature. The ground state is then the minimum with the lowest energy. 10 systems that are found non-magnetic at this step are discarded.
    }
\end{figure*}
\section*{Results and Discussions}
\label{sec3}

To identify magnetic 2D materials and their magnetic ground state, we consider monolayers from MC2D,\cite{campi_materials_2022,mc2d} a portfolio of 2D materials that can be exfoliated from experimentally known 3D parent compounds, identified using high-throughput computational exfoliation.~\cite{mounet_two-dimensional_2018,campi_expansion_2023} We start from 877 easily exfoliable 2D materials with up to 12 atoms per unit cell. As summarized in~\cref{workflow}, the tendency to show magnetism is first tested using an improved version of the AiiDA~\cite{pizzi_aiida_2016, huber_aiida_2020} workflow, named Chronos, originally adopted in Ref.~\citenum{mounet_two-dimensional_2018} to screen a smaller set of materials. For each system, the workflow first considers several ferromagnetic configurations through different starting magnetizations on the atoms and computes its optimized geometry and total energy through collinear DFT calculations with an ordinary approximate exchange-correlation functional (PBE). If any ferromagnetic state is found to have a total energy lower than a non-magnetic reference calculation, the system is considered potentially magnetic and additional magnetic configurations are investigated, including antiferromagnetic states that require up to a $2\times1$ supercell to accommodate two magnetic atoms with opposite spins (see Methods for more details). In this way, 228 magnetic monolayers are identified with up to 12 atoms in their primitive cell, together with their potential collinear ground state. 
Out of these, 174 are found to be ferromagnetic, and further broken down into 100 metals and 74 insulators. The remaining 54 materials show an AFM ground state, with 31 having a finite gap and 23 being metallic.\\

Although versatile and flexible, the workflow explores possible magnetic states only through the starting magnetization, that is, through initial conditions on the density of spin up and spin down states on each atom. A more robust way is to initialize from the atomic orbital occupation matrices - particularly $d$ and $f$ orbitals where the magnetization usually stems from - and explore more possible states.
Introducing atomic orbitals $\varphi_{m_1}^I(\mathbf{r})$ labeled by $m$ on the $I$-th atom, we can define an atomic occupation matrix as $n^{I\sigma}_{m_1m_2}~=~\sum_{v\mathbf{k}} f_{v\mathbf{k}\sigma} \langle \psi_{v\mathbf{k}\sigma} | \varphi_{m_2}^J \rangle \langle \varphi_{m_1}^I | \psi_{v\mathbf{k}\sigma} \rangle$, where $f_{v\mathbf{k}\sigma}$ are the occupations of Bloch states $|\psi_{v\mathbf{k}\sigma} \rangle$ and magnetization is an unbalance between the occupation of spin-up and down states $m^I~=~\sum_{m} \left( n^{I\uparrow}_{m m} - n^{I\downarrow}_{m m} \right)$.
To explore more systematically the magnetic energy landscape and obtain a more reliable identification of the magnetic ground state, it is important to gain full control over the atomic occupation matrix (at least for the relevant atomic orbitals on the magnetic atoms), not just through the spin density.~\cite{payne_optimizing_2019,allen_occupation_2014,dorado_stability_2010,dorado_dft_2009,meredig_method_2010} We thus validate the magnetic ground-state identification by applying a constraint on the atomic orbital occupations and then letting the self-consistent procedure find the closest local minimum in the energy landscape. The overall global minimum corresponding to the magnetic ground state can then be found by exhaustively exploring possible atomic occupations and comparing the energies of the corresponding self-consistent solutions. To perform this task automatically and efficiently, we use the RomeoDFT workflow~\cite{ponet_energy_2024} (see Methods for more details). We first consider a subgroup of magnetic monolayers and find that without Hubbard corrections in essentially all cases the same ground state is found either by controlling the starting magnetization by spin density or the constraints on the full atomic occupation matrix (see figure S1 in Supporting Information. When considering PBE calculations, the search implemented in the Chronos workflow is thus exhaustive enough to reproduce the more sophisticated exploration performed by RomeoDFT. 

\begin{figure}[htbp]
    \centering
    \captionsetup{justification=Justified,singlelinecheck=false}
    \includegraphics[width=0.8\linewidth]{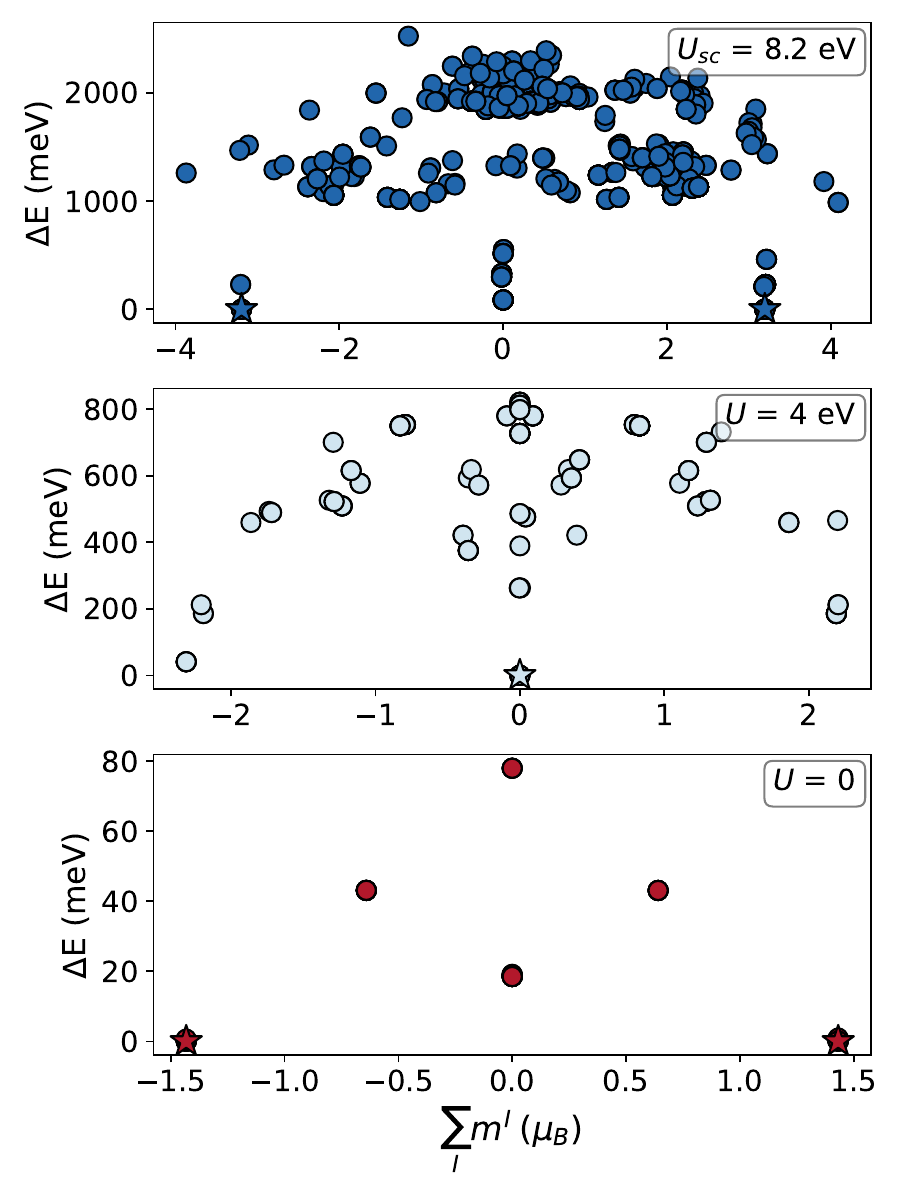}
    \caption{Self-consistent states for monolayer CoO$_2$ from PBE+$U$ calculations with different values of $U$ with the star denoting the ground state identified by the lowest energy. There are six atoms in the unit cell to accommodate AFM ordering. $U_{sc}$ shows the $U$ value that is calculated by DFPT. At the PBE level ($U=0$) and at PBE+$U_{sc}$ the system is ferromagnetic with different values of magnetic moments. For an intermediate value of $U$ at 4 eV the system exhibits antiferromagnetism. Therefore, selecting the correct $U$ value is crucial to find the correct magnetic ordering, and using a method to explore the energy landscape can help avoid converging to an incorrect state.
    At $U=0$, the number of states is smaller, which means simpler methods such as Chronos can still effectively find the ground state (see Supporting Information for more details on the case with $U$ = 0).}
    \label{O2Co}
\end{figure}

Most of the identified 2D magnetic materials host electrons in localized $d$ or $f$ orbitals, for which approximate DFT functionals such as PBE tend to suffer from inaccuracies arising from self-interaction.~\cite{kulik_density_2006,cohen_insights_2008} The magnetic ground-state predictions might be affected by such errors and need to be validated against more accurate calculations. As alluded to before, an approach to mitigate self-interaction problems, known as DFT+$U$, relies on complementing DFT functionals with terms inspired by the Hubbard model.~\cite{bao_self-interaction_2018, himmetoglu_hubbard-corrected_2014, mosquera_derivative_2014, anisimov_band_1991, anisimov_density-functional_1993, solovyev_corrected_1994, liechtenstein_density-functional_1995, dudarev_electron-energy-loss_1998, perdew_density-functional_1982} Such Hubbard-corrected functionals have been shown to remove prominent self-interaction errors when the Hubbard $U$ parameter is computed self-consistently from linear-response theory,~\cite{cococcioni_linear_2005} thus preserving the parameter-free character of DFT. Unfortunately, the inclusion of Hubbard corrections makes the energy landscape much more complex, with the appearance of many emergent local minima with similar total energies.~\cite{ponet_energy_2024} To illustrate this effect, we consider the case of monolayer CoO$_2$ in~\cref{O2Co}, where each point represents a local minimum in the energy landscape, identified by exploring multiple occupation matrices. At the PBE level ($U = 0$), only a limited set of local minima is found, with an overall ferromagnetic ground state (indicated by a star). When including Hubbard corrections, with the Hubbard $U$ calculated self-consistently from the linear-response method (PBE+$U_{\rm sc}$), the number of local minima significantly increases, making the magnetic energy landscape more complex and the possibility of falling in a local minimum different from the true ground state higher. We thus have that, while at the PBE level ($U=0$) the number of states is limited and the identification of the global ground-state minimum is simpler (explaining the agreement between Chronos and RomeoDFT at the PBE level --- see fig.~S1 in the Supporting information), an approach based on the occupation matrix is mandatory when Hubbard corrections are included. In~\cref{O2Co} we also show results for an intermediate case with $U=4$~eV, where the proliferation of local minima is already present, although less dramatic than at $U_{\rm sc}$. Remarkably, the ground state, labeled by a star in each panel, is very sensitive to the value of $U$ as the system is predicted to be ferromagnetic at $U=0$, antiferromagnetic at 4~eV and then ferromagnetic again (but with a different magnetization) at $U_{\rm sc} = 8.2$~eV.~\cite{liang_tunable_2021} Therefore, the choice of $U$  strongly affects the ground-state properties of the system. To avoid any empiricism, we compute the Hubbard parameter $U$ for each material using an efficient formulation of the linear-response approach~\cite{cococcioni_linear_2005} within density-functional perturbation theory \cite{timrov_self-consistent_2021, timrov_hp_2022, timrov_hubbard_2018} (see Methods and Supporting Information for more details).

\begin{figure*} \centering\captionsetup{justification=Justified,singlelinecheck=false}
    \includegraphics[width=\textwidth]{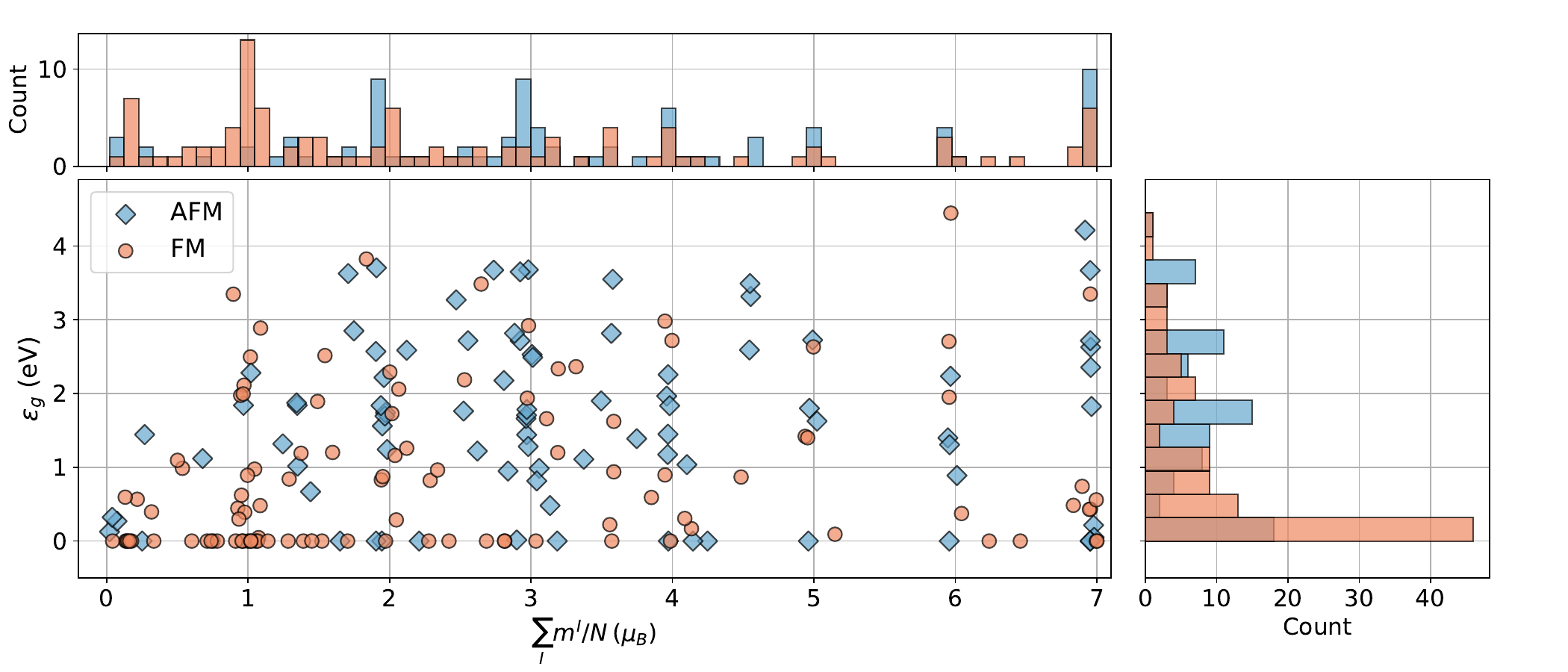}
    \caption{Band gap value, density of magnetic moments per atom, and their distribution for the ground state of the magnetic monolayers studied in this work obtained by exploring their energy landscape with PBE+$U$ calculations. $N$ is the number of magnetic atoms per cell, I is the index of the magnetic atoms and $m^I$ is the magnetization defined from occupation matrix ($m^I~=~\sum_{m} \left( n^{I\uparrow}_{m m} - n^{I\downarrow}_{m m} \right)$, for more details see the methods section). Orange and blue show ferromagnetic and antiferromagnetic systems, respectively.}
    \label{total}
\end{figure*}

Equipped with self-consistent Hubbard corrections, we navigate the magnetic energy landscape of potentially 2D magnetic materials identified through the Chronos workflow at the PBE level, now using PBE+$U$ and a control over the atomic occupation matrix using RomeoDFT. Out of 228 materials, 34 systems are discarded because either they are found to be nonmagnetic or there are convergence issues in the calculation of the Hubbard parameter or self-consistent calculations, leaving us with 194 magnetic monolayers for which we characterize their ground state. The results are summarized in~\cref{total}, where we report the absolute magnetization per magnetic atom and the band gap of the ground state of all systems, together with their distribution. More details on the specific magnetic configuration, crystal structure, magnetic energy landscape, and band structure for each material are provided in the Supporting Information.

\begin{figure}[h]
    \centering
    \includegraphics[width=0.5\textwidth]{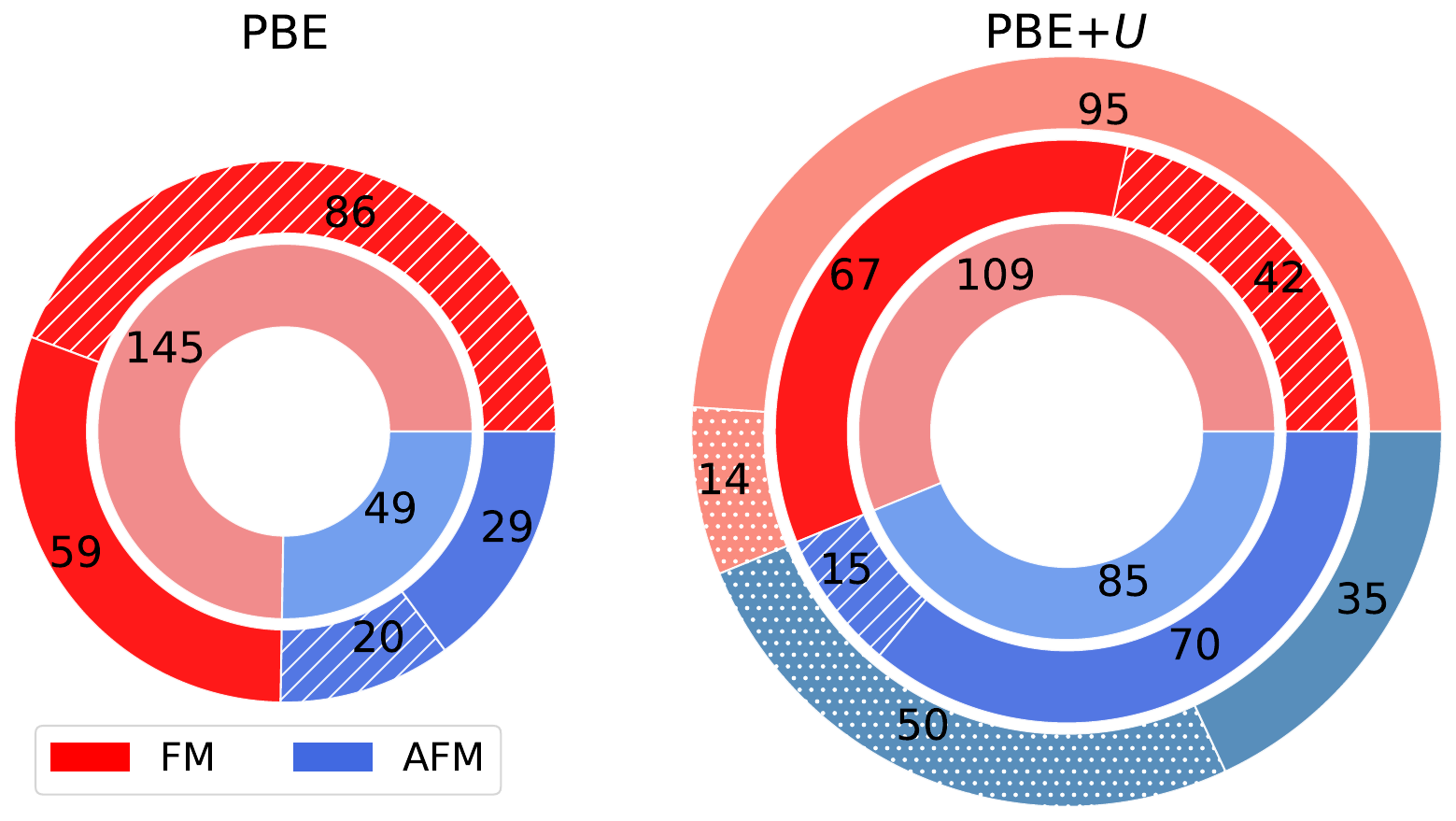}
    \caption{Breakdown of magnetic monolayers identified by two approaches: PBE calculations using Chronos workflow on the left and PBE+$U$ calculations using RomeoDFT workflow on the right (see~\cref{workflow}). Blue indicates ferromagnetic systems, and red represents antiferromagnetic systems. The inner rings display the number of materials within each magnetic category. The second rings differentiate between metallic (dashed fill) and semiconducting (solid fill) systems for each category. The outermost ring on the right diagram indicates the number of systems that either remained in the same magnetic category as identified by the Chronos workflow (solid fill) or shifted categories (dotted fill). Note that the two altermagnets are considered in the antiferromagnetic category.}
    \label{pie}
\end{figure}

We categorize all systems into four groups, depending on whether they are ferromagnetic or antiferromagnetic, insulators or metals (we did not find ferrimagnets, and the two altermagnets are included in the antiferromagnetic group).
\Cref{pie} shows the number of systems in each group, with the left and right charts representing results from PBE (Chronos workflow) and PBE+$U$ (RomeoDFT workflow) calculations, respectively. Blue and red colors indicate antiferromagnetic and ferromagnetic materials. The innermost rings show the number of systems in each category, while the second rings count metallic (hatched) or semiconducting (solid) systems. In the right chart, the outermost ring indicates the number of systems from the innermost ring that either remained in the same magnetic category (solid fill) or changed categories (dotted fill) from the PBE calculations. As expected, the inclusion of Hubbard corrections leads to an increase in the number of insulating systems. From \cref{pie}, we also note that using the RomeoDFT workflow with PBE+$U$ calculations significantly expands the number of antiferromagnetic systems (blue colors): approximately 40\% of these materials were identified as ferromagnetic when using the Chronos workflow with standard PBE calculations. This change can be attributed primarily to two factors: first, the limited capability to explore the energy landscape when using simply a starting magnetization; second, the possible increase in the number of magnetic states after including Hubbard corrections.\\

 We note in passing that among the AFM identified above, we find only two altermagnetic monolayers by generalizing the protocol for 3D systems~\cite{smolyanyuk_tool_2024} to include additional symmetry constraints in 2D.~\cite{sodequist_two-dimensional_2024,zeng_description_2024} These are RuF$_4$ and VF$_4$, with the same structural prototype. Both systems were already predicted to be 2D altermagnets in Ref.~\citenum{sodequist_two-dimensional_2024}; we do not see altermagnetism in FeBr$_3$, because its crystal structure and symmetry is different from the one studied in Ref.~\citenum{sodequist_two-dimensional_2024}.

Due to the scarcity of experimental results for magnetic monolayers, it is challenging to validate our predictions against experiments. We thus focus on a few cases reported in the existing literature. For example, our results agree with the experimentally observed ferromagnetism in the family of chromium trihalides (CrX$_3$; X~=~Cl, Br, I with CrCl$_3$ being 2D-XY ferromagnetic)~\cite{chen_direct_2019, huang_layer-dependent_2017,bedoya-pinto_intrinsic_2021} and in CrSBr.~\cite{lee_magnetic_2021}
In the family of transition-metal phosphorus trisulfides, our results match the experimentally observed ferromagnetism in CrPS$_4$~\cite{son_air-stable_2021} and the antiferromagnetic ordering in NiPS$_3$, although suppressed by fluctuations in the monolayer limit.~\cite{kim_suppression_2019} For the case of FePS$_3$, our findings show ferromagnetic ordering, in contrast with the antiferromagnetic ordering observed in experiments.~\cite{lee_ising-type_2016} We attribute this discrepancy to the limited size of the unit cell considered in this work for FePS$_3$, which is not able to accommodate the measured zigzag ordering. Indeed, since the primitive cell already contains two magnetic atoms, no supercell is considered, and only a N\'eel AFM ordering is tested. This limitation, combined with the use of collinear calculations, could influence the accuracy in predicting other antiferromagnetic systems and may lead to discrepancies with other studies,~\cite{olsen_antiferromagnetism_2024,sodequist_magnetic_2024} and we expect it to be at the origin of the relatively small number of AFM systems with respect to FM.\\

\begin{figure*}[t] \centering\captionsetup{justification=Justified,singlelinecheck=false}
    \includegraphics[width=\textwidth]{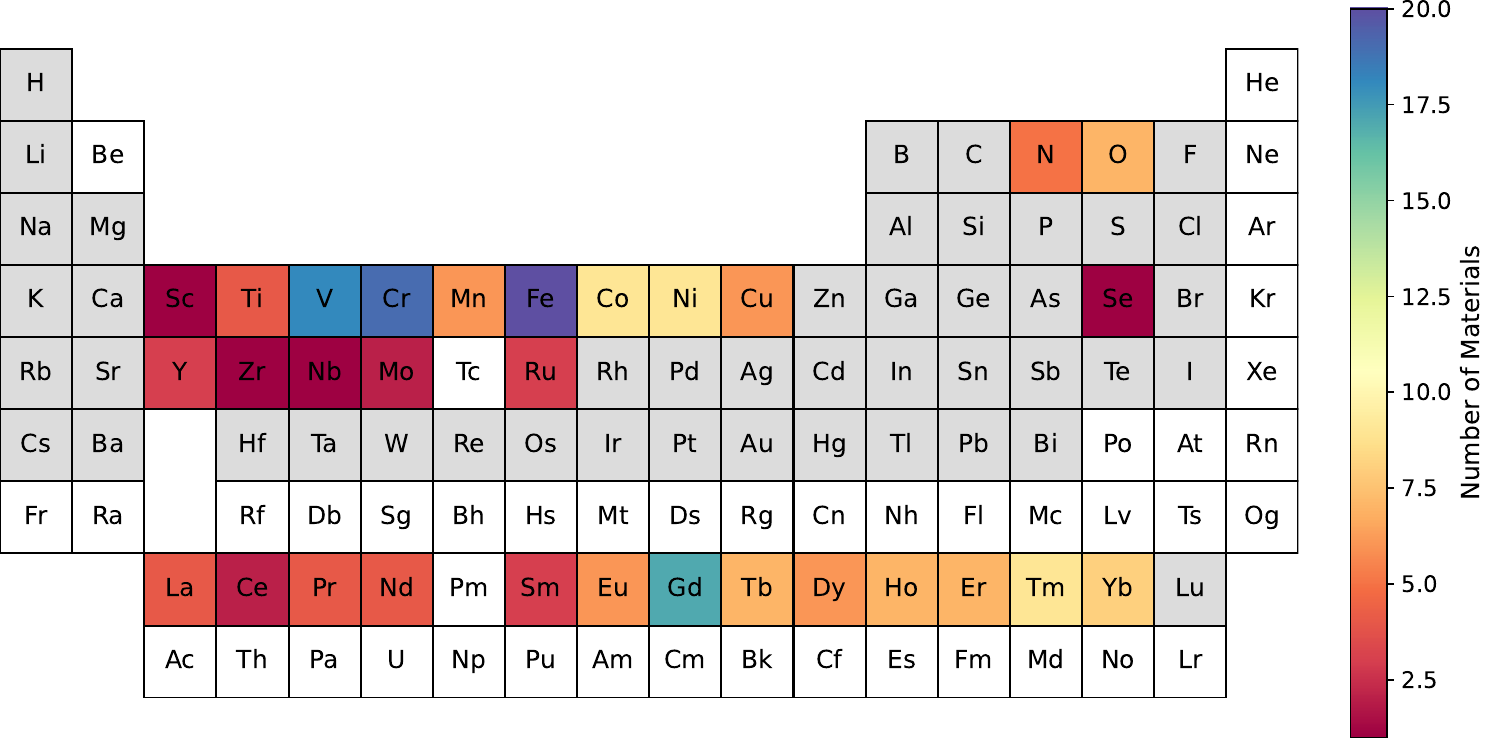}
    \caption{ Periodic table displaying color intensity to indicate the number of magnetic monolayers in which each atom is magnetic. An atom is considered magnetic in a system if its magnetic moment exceeds 0.1 $\mu_B$ and it either (i) has the highest magnetic moment in the system or (ii) possesses at least 80\% of the maximum magnetic moment in the system. Atoms with a white background are not present in the initial set of 877 systems considered
in the Chronos workflow.}
    \label{heatmap}
\end{figure*}

Among the possible conclusions that we can draw from the high-throughput investigation of the magnetic ground-state of 2D materials, we also focus on the nature of magnetic elements in each structure. Conventional magnetism usually arises from $d$ or $f$ orbitals, qualifying transition metals and lanthanides as the dominant contributors to magnetism. Our results largely confirm such conventional wisdom, as shown in~\cref{heatmap}, where we report the number of systems in which a given element is magnetic. However,~\cref{heatmap} also suggests that certain chalcogen and nonmetal elements (especially oxygen) also play a role in magnetism in 2D compounds, putting forward their potential role in unconventional magnetism. For instance, in GaSe (space group: $p\Bar{3}m1$) and SiN$_2$F$_6$, the selenium and nitrogen atoms exhibit non-negligible magnetic moments (0.2~$\mu_B$ and 2~$\mu_B$ respectively). Interestingly, in the case of CdOCl, while the $d$ orbitals of Cd are fully occupied and exhibit negligible magnetic moments, magnetism predominantly arises from the oxygen atoms, which have a magnetic moment of 0.8~$\mu_B$.\\

 Exploring the energy landscape gives us access not only to the ground state of the system but also to the ability to quantify the energy difference between the ground state and other metastable magnetic states. This is particularly relevant when it is possible to identify a metastable state with magnetic moments on the atoms similar to the ground state but different magnetic ordering, \textit{e.g.}, a ferromagnetic ground state and a metastable antiferromagnetic state with comparable absolute magnetization (or vice-versa). The energy difference between these states can then be translated into an effective exchange parameter $\tilde J$ of an underlying  Heisenberg model through
\begin{equation}
    \tilde J~=~\frac{1}{2N} (E_{FM}-E_{AFM})
    \label{heisenberg}
\end{equation}
 where $N$ is the number of magnetic atoms per cell while $E_{FM}$ and $E_{AFM}$ are the energies of the ferromagnetic and antiferromagnetic states, respectively. This effective exchange parameter $\tilde J$ is related to an effective nearest-neighbor exchange coupling $J$ through $\tilde J = n J$, where $n$ is the number of nearest neighbors. A large energy difference, $|E_{FM}-E_{AFM}|$, corresponds to a large exchange parameter $\tilde J$,  stabilizing the magnetic order and possibly resulting in a higher critical temperature. It is important to note that the mapping between first-principles calculations and the Heisenberg model is meaningful only when the magnetization arises from local atomic magnetic moments that have the same values (although different orientations) in the FM and AFM states, \textit{i.e.}, $\mid$S$_{FM}$$\mid$~=~$\mid$S$_{AFM}$$\mid$. This condition is typically met in insulators, while in general it cannot be readily applied to magnetic metals, where magnetism arises from itinerant electrons. Therefore, we consider only magnetic insulators to compute $E_{FM}-E_{AFM}$ and the effective $\tilde J$. \\

\begin{figure*}
    \centering  \captionsetup{justification=Justified,singlelinecheck=false}
    \includegraphics[width=1\textwidth]{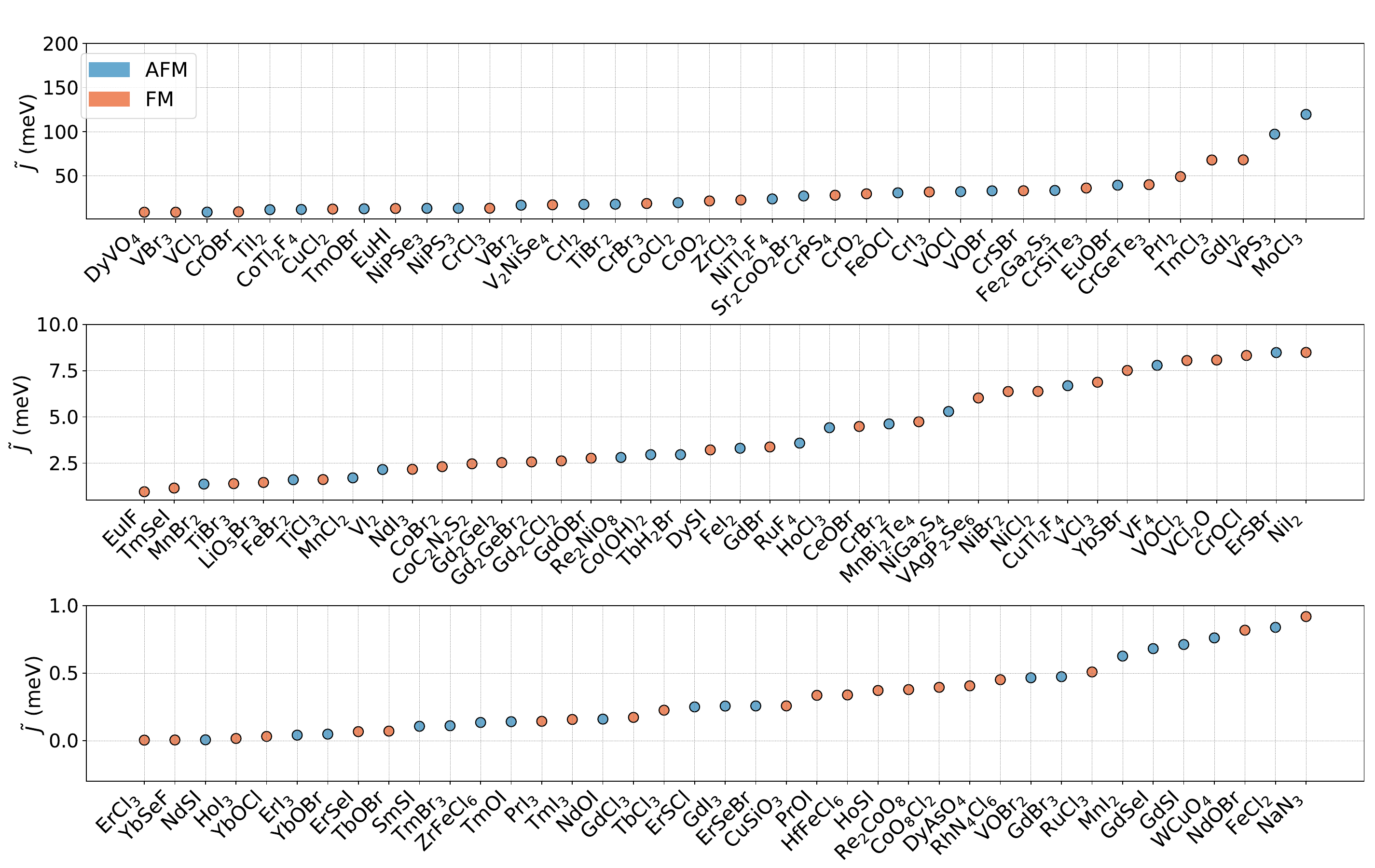}
    \caption{ Effective exchange parameter, $\tilde J$, calculated from~\cref{heisenberg} using the energy difference between the ground state and the closest state with opposite magnetic configuration (with the condition of $|S_{FM}|~=~|S_{AFM}|$) obtained by exploring the energy landscape. Blue and orange circles represent antiferromagnetic and ferromagnetic ground states, respectively. The $y$-axis range progressively increases across the three subplots.}
     \label{energy_difference}
\end{figure*}

\Cref{energy_difference} shows $|E_{FM}-E_{AFM}|$ divided by the number of magnetic atoms per cell, $i.e.$ $\tilde J$, for the insulating magnetic monolayers for which it was possible to find a metastable magnetic state with similar atomic magnetic moments but different ordering. Systems are sorted with increasing effective $\tilde J$. The bottom panel contains systems where exchange interactions are weak, making it more challenging for the magnetic ordering to compete with thermal fluctuations. On the contrary, the systems in the top panel are expected to have higher critical transition temperatures. The ranking based on the effective $\tilde J$ seems to reproduce several experimental trends for the ordering temperature. For instance, it is experimentally known that CrSBr has a higher critical temperature than the chromium trihalides,~\cite{huang_layer-dependent_2017, lee_magnetic_2021} and within this family the iodine compound has the largest $T_c$ while the chlorine one the smallest,~\cite{kim_micromagnetometry_2019, kim_evolution_2019, zhang_direct_2019, huang_layer-dependent_2017,bedoya-pinto_intrinsic_2021} in agreement with the ordering in~\cref{energy_difference}.
 
To further validate $\tilde J$ as a meaningful proxy for $T_c$, we select three systems with the largest \( J \) in~\cref{energy_difference} and compute their transition temperatures using different approaches. We thus focus on MoCl$_3$, GdI$_2$, and VPS$_3$, noting that the latter was already identified as a potentially high-$T_c$ magnetic monolayer based on high-throughput calculations.~\cite{torelli_high-throughput_2020} Additionally, we include CrI\(_3\) in our analysis, as experimental results are available for comparison. In addition to the approach based on~\cref{heisenberg} and relying on the energy difference between FM and AFM reference states, we also consider the so-called four-state method,~\cite{xiang_predicting_2011} which calculates the exchange parameter based on the energy differences between four distinct magnetic configurations of a pair of neighboring atoms in a supercell (see Methods for more details). Remarkably, the four-state method allows the calculation of specific interatomic exchange parameters, not just an effective nearest-neighbor $J = \tilde J/n$ as in~\cref{heisenberg}, and we consider interactions up to the third (fourth) nearest neighbors for CrI$_3$ and GdI$_2$ (VPS$_3$ and MoCl$_3$). The exchange parameters computed within the two approaches are then denoted \( J^{FM-AFM} \) and  \( J^{4states} \) in the following. From the knowledge of the exchange parameters, the transition temperature is calculated using either mean-field (MF) theory or classical Monte Carlo (MC) simulations. In the latter, we completely neglect relativistic effects. Indeed, magnetocrystalline anisotropy does not affect significantly the critical temperature for a finite-size system, and even isotropic monolayers can sustain magnetic order at finite temperature when considering realistic sample sizes.~\cite{jenkins_breaking_2022}

\begin{figure}[h]
    \centering    \captionsetup{justification=Justified,singlelinecheck=false}
    \includegraphics[width=0.5\textwidth]{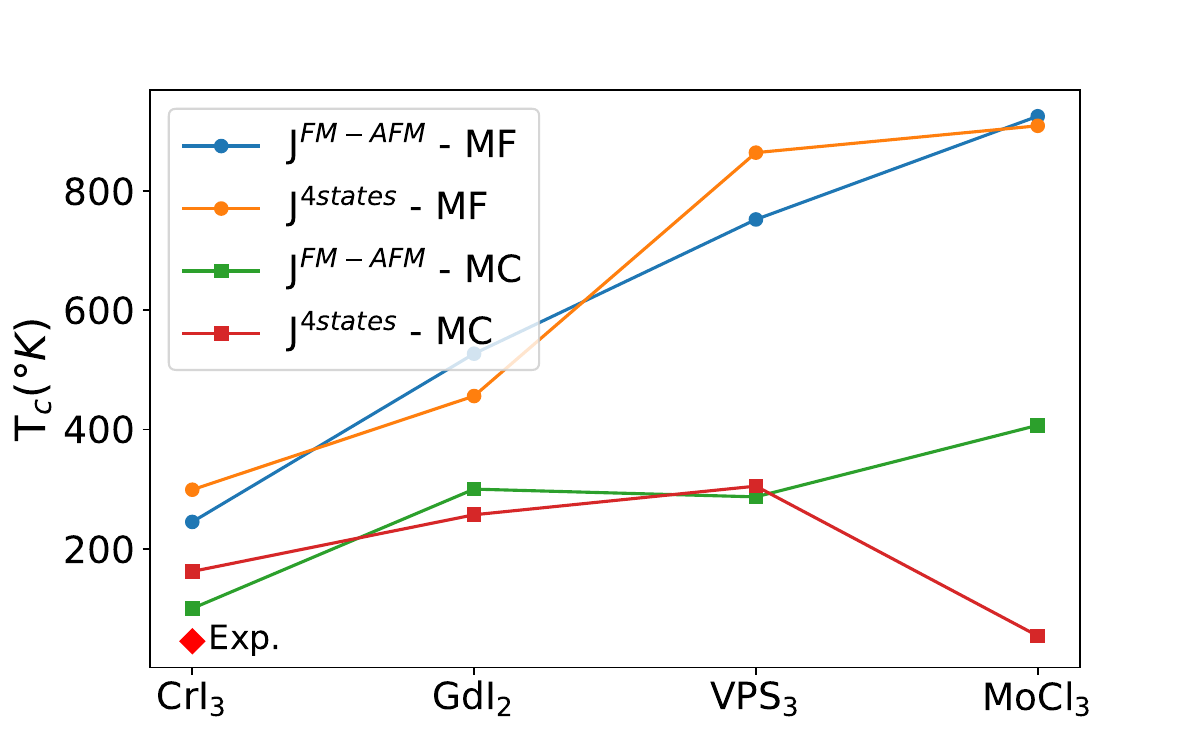}
    \caption{The transition temperature of three magnetic monolayers (GdI$_2$, VPS$_3$, and MoCl$_3$) with the highest ~$\tilde J$ in~\cref{energy_difference} from two methods: mean-field (MF) and Monte Carlo (MC) calculations. The exchange parameters used in the calculation of T$_c$ are obtained using two different approaches: 1- exploring the energy landscape and calculating the energy difference of the FM and AFM states ($J^{FM-AFM}$) as given in~\cref{heisenberg}, and 2- four-states method ($J^{4states}$).}   
    \label{tc}
\end{figure}

Critical temperatures computed either using MF or MC, with exchange parameters estimated from both \cref{heisenberg} and the four-state method, are shown in~\cref{tc} for CrI$_3$, GdI$_2$, VPS$_3$, and MoCl$_3$. Trends in \( T_c \) derived from both MC and MF methods are consistent with \cref{energy_difference}, with MF calculations showing higher values, as expected from the well-known overestimation of $T_c$ in mean-field theory.~\cite{garanin_self-consistent_1996} 
We note that, in general, different methods for computing the exchange parameters give similar values of \( T_c \). The only exception is MoCl\(_3\), for which the MC estimates of \( T_c \) differ significantly between the two methods, with a notably lower value when the exchange parameters are calculated using the four-state method. This discrepancy is attributed to the dimerization of Mo atoms in MoCl\(_3\), which distorts the otherwise honeycomb lattice of Mo atoms. The energy difference between the FM and AFM is mainly arising from a strong AFM coupling between dimerized pairs, while dimers are only weakly coupled. This feature is correctly captured within the four-state method, while the effective $\tilde J$ approach of \cref{heisenberg} uniformly distributes the FM-AFM energy difference between the $n=3$ closest atoms. This failure also affects all MF estimates in a similar way. We thus expect that the effective $\tilde J$ reported in \cref{energy_difference} provides a good estimation of the ranking in critical temperature, provided that exchange interactions are distributed evenly between similar neighbors.\\

We also note that the transition temperature of the CrI$_3$ monolayer is higher than the values reported in previous studies and experiments\cite{pizzochero_magnetic_2020,torelli_first_2020,huang_layer-dependent_2017}. This discrepancy can be attributed to differences in geometry, the relatively large $U$ value for CrI$_3$ obtained from linear-response,\cite{haddadi_-site_2024} and the influence of broadening effects in Brillouin-zone integrations. For consistency, we employed the same broadening value as used in the Chronos workflow (270~meV). 
Since the exchange parameter is derived from total energies--which are sensitive to the choice of broadening--it is likewise influenced by the selected broadening value.

\begin{figure*}[t]
    \centering    \captionsetup{justification=Justified,singlelinecheck=false}
    \includegraphics[width=1\textwidth]{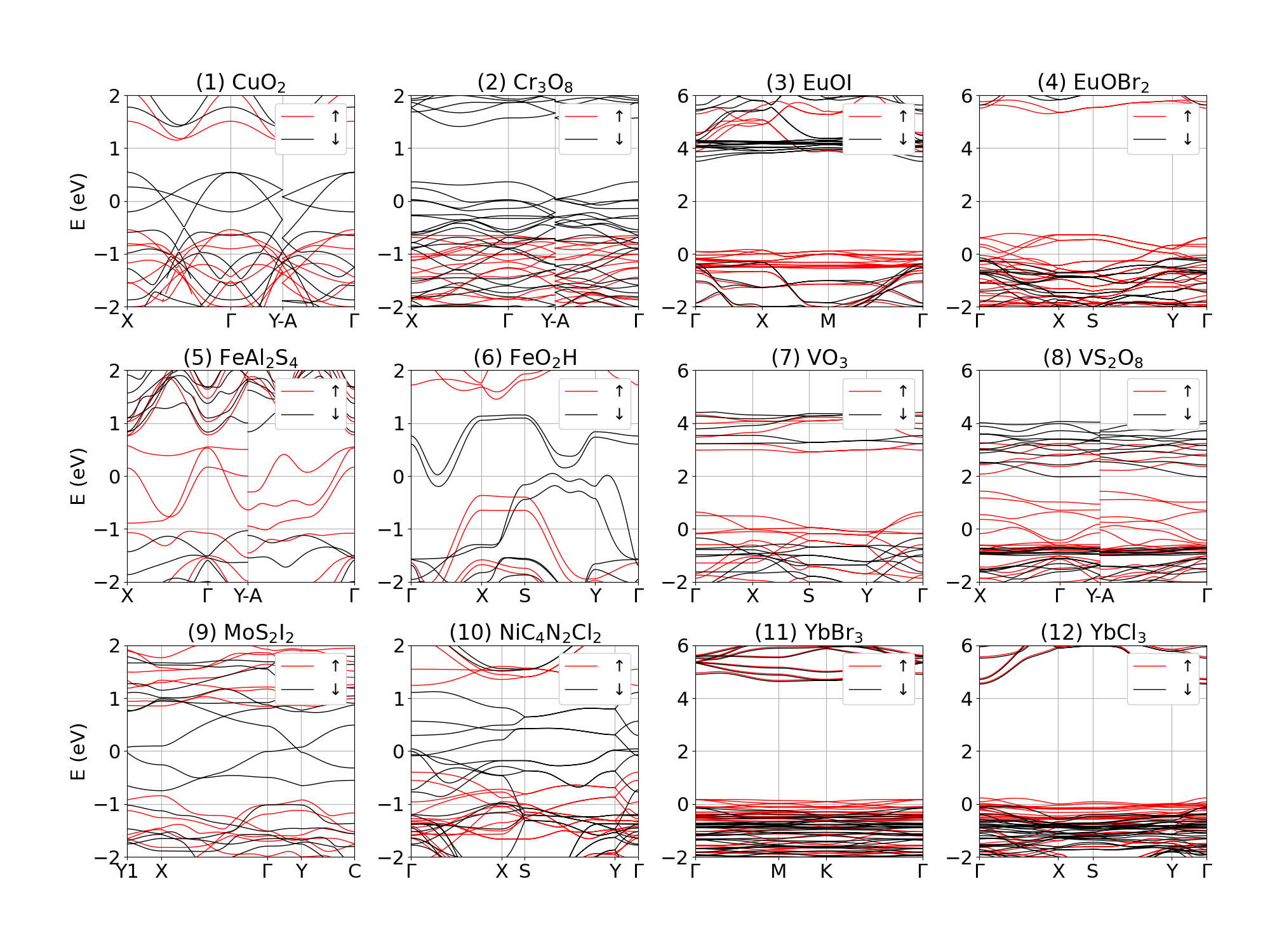}
    \caption{Electronic bands of half metals identified in this work. The Fermi energy is set at zero. Note that the two rightmost columns have a larger energy window to be able to show the band gap of the opposite spin channel.}
    \label{half_metals}
\end{figure*}

Among metallic systems, we are particularly interested in half-metallic ferromagnetic systems that are interesting for spintronicss devices because they can provide fully spin-polarized currents.~\cite{li_first-principles_2016} We thus search for systems where the Fermi energy crosses the energy bands for one spin channel while the other spin channel remains gapped.~\cite{de_groot_new_1983} Twelve novel half-metals are identified, whose electronic structure is reported in~\cref{half_metals}. In several cases the bandwidth is sufficiently large to expect spin-polarized transport even in the presence of disorder. Remarkably, in systems like CuO$_2$, Cr$_3$O$_8$, EuOI, EuOBr$_2$, VO$_3$, VS$_2$O$_8$, YbBr$_3$, and YbCl$_3$ a large gap is present above the metallic spin channel, which is potentially interesting in view of possible optoelectronic control of spin currents. In all these cases a correct description of the magnetic ground state is crucial to have half-metallic properties. For instance, in the case of FeAl$_2$S$_4$ it is important to have the two iron atoms with different oxidation states (+1 and +2) and thus different magnetic moments of 2.2~$\mu_B$ and 3.2~$\mu_B$. We also stress that, in general, the half-metallic state can be fragile, especially if it is protected by symmetry-induced degeneracies.~\cite{yao_fragile_2021} Further tests to include the effect of spin-orbit coupling or structural distortions would thus be needed to fully validate predictions, although the procedure outlined above is already sufficient to discard speculative half-metals such as iron dihalides FeX$_2$ (with X~=~Cl, Br, I). These systems have been predicted to be half-metals,~\cite{ashton_two-dimensional_2017, torun_stable_2015} although experimental evidence~\cite{hadjadj_epitaxial_2023, prayitno_controlling_2021, zhou_evidence_2024, kong_magnetic_2020,cai_fecl2_2020}suggests that they have a finite gap. We find them to be indeed insulating, although the mechanism to break the symmetry-induced degeneracy arises from an incorrect AFM configuration, as we do not include relativistic effects or distortions.

\section{Conclusions}
In conclusion, we investigate the magnetic ground states of easily exfoliable monolayers from the MC2D database. Starting from 3077 easily and potentially exfoliable monolayers identified in Refs.~\citenum{campi_expansion_2023, mounet_two-dimensional_2018}, an initial screening is carried out on 877 easily-exfoliable systems containing up to 12 atoms per cell using high-throughput density functional theory with the conventional PBE exchange-correlation functional. Testing various magnetic configurations by changing the initial magnetization, 228 magnetic monolayers are identified as potentially magnetic systems. Subsequently, Hubbard corrections are introduced for these systems, with the Hubbard parameter determined self-consistently. The magnetic energy landscape is then thoroughly explored by applying constraints to the occupation matrices of the $d$ or $f$ orbitals, enabling the identification of the global energy minimum.
Through this process, we identify 194 magnetic monolayers, comprising 109 ferromagnetic, 83 antiferromagnetic, and 2 altermagnetic systems. To estimate critical temperatures and highlight promising candidates, we employ a proxy based on the energy difference between ferromagnetic and antiferromagnetic states. This method is applied to the 137 monolayers that exhibit an electronic band gap. Among the remaining 57 metallic systems, we identify 12 novel half-metals with strong potential for spintronics applications.
\label{sec4}

\section*{Methods}
\label{methods}
In this work, all calculations are collinear and performed using the \textsc{Quantum ESPRESSO} (QE) distribution.~\cite{giannozzi_quantum_2009,giannozzi_advanced_2017,giannozzi_q_2020} Calculations are performed within the
spin-polarized generalized-gradient approximation (GGA) for exchange-correlation functional using the Perdew-Burke-Ernzerhof (PBE) prescription. The pseudopotentials and the wavefunction kinetic-energy cutoffs are taken from a standard solid-state pseudopotentials (SSSP Efficiency) library.~\cite{prandini_precision_2018,lejaeghere_reproducibility_2016,vanderbilt_soft_1990,dal_corso_pseudopotentials_2014,garrity_pseudopotentials_2014} We use the most recent versions of the SSSP library available at the time for our calculations. Specifically, version 1.1.2 is used for the PBE calculations, and version 1.3.0 is utilized for the PBE+$U$ calculations. The $\mathbf{K}$-points density of $0.2~{\rm\AA}^{-1}$ and $\mathbf{q}$-points mesh density of $0.4~{\rm\AA}^{-1}$ have been set for single self-consistent (scf) and density functional perturbation theory (DFPT) calculations, respectively. In this work, a system (an atom) is magnetic if the absolute total magnetization per cell (of the atom) is larger than 0.1$~\mu_B$. The data used to produce some of the results of this paper are given in the materials cloud archive.~\cite{materialscloud_arxiv}\\

In this work we consider systems from MC2D,~\cite{campi_materials_2022, mc2d,campi_expansion_2023} which is created from a collection of 3077 exfoliable monolayer that were identified from prototypes of the Inorganic Crystal Structure Database (ICSD),~\cite{noauthor_inorganic_nodate, bergerhoff_inorganic_1983} the Crystallography Open Database (COD),~\cite{grazulis_crystallography_2012} and Materials Platform for Data Science (MPDS)~\cite{noauthor_materials_nodate} databases. We identify and characterize magnetic monolayers in MC2D in two main steps, as shown in~\cref{workflow}: in the first step (Chronos step), using PBE calculations, potentially magnetic monolayers are identified, and in the second step (RomeoDFT step), after including Hubbard corrections and exploring the energy landscape, the magnetic ground state of these materials is found. An initial screening was performed using an automated AiiDA~\cite{pizzi_aiida_2016, huber_aiida_2020} workflow (Chronos) on 877 easily exfoliable materials (\textit{i.e.} materials with a binding energy per unit of area lower or equal to 30 meV/\AA$^{2}$ when computed with the vdW-DF2-c09 functional or lower or equal to 35 meV/\AA$^{2}$ when computed with the rvv10) with up to 12 atoms per unit cell, using PBE calculations, and  228 magnetic monolayers are identified by the following procedure: the geometry of the system is optimized in ferromagnetic and non-magnetic orderings. If the ferromagnetic ordering shows lower energy than the non-magnetic case, the system is considered to be potentially magnetic, and other antiferromagnetic orderings (making sure the supercell size is now $2\times1$ to accommodate two magnetic atoms) and five random configurations are created. The system is optimized starting from different starting magnetization defined by spin density ($m^I = \int (\rho^{I,\uparrow} (\textbf{r}) - \rho^{I,\downarrow} (\textbf{r})) d\textbf{r}$ where $\rho$ describes the density and I is the atom index). As stated before this simplistic approach is effective in capturing the correct magnetic ground state in the absence of Hubbard correction (see Supporting Information). However, it is important to note that there are still local minima within the energy landscape that are accessible through the manipulations of the orbital occupations.

This can be achieved using RomeoDFT, which corresponds to the second step of our overall workflow. In this approach, the system is induced towards a set of target occupation matrices $\Tilde n^I$ using Lagrange multipliers:
\begin{equation*}
    \Tilde{E} = E_{DFT} + E_{U} + \sum_{I, m_1 , m_2} \lambda^I_{\alpha, \beta} (n^I_{m_1 , m_2} - \Tilde{n}^I_{m_1 , m_2}) \,
\end{equation*}
with
\begin{equation}
    E_{U} =  \frac{1}{2} \sum_{I} \sum_{\sigma m_1m_2} U^I (\delta_{m_1m_2} - n^{I\sigma}n^{I\sigma}_{m_1m_2}) n^{I\sigma}_{m_2m_1} \,
    \label{eqn2}
\end{equation}
where E$_U$ is the Hubbard corrected energy, $\lambda$ is the Lagrange multiplier, representing the strength of the energy penalty associated with a deviation from the target occupation matrices, $I$ is the atomic site index, $m_1$ and $m_2$ are the magnetic quantum numbers associated with a specific angular momentum and $U^I$ is the on-site Hubbard parameters. $n^{I\sigma}_{m_1m_2}$ are the generalized atomic occupation matrices which are computed by projecting the Kohn-Sham (KS) wavefunctions $\psi_{v\mathbf{k}\sigma}(\mathbf{r})$ on atomic orbitals $\varphi_{m_1}^I(\mathbf{r})$ as: $n^{I\sigma}_{m_1m_2} = \sum_{v\mathbf{k}} f_{v\mathbf{k}\sigma} \langle \psi_{v\mathbf{k}\sigma} | \varphi_{m_2}^J \rangle \langle \varphi_{m_1}^I | \psi_{v\mathbf{k}\sigma} \rangle$, where $f_{v\mathbf{k}\sigma}$ are the occupations of KS states. The general formulation of DFT+$U$ is discussed in Refs.~\citenum{dudarev_electron-energy-loss_1998,leiria_campo_jr_extended_2010, timrov_hubbard_2018,timrov_self-consistent_2021}. We note that in this case the magnetization is defined from the occupation matrix of spin up and down: $m^I~=~\sum_{m} \left( n^{I\uparrow}_{m m} - n^{I\downarrow}_{m m} \right)$.
After imposing the constraint for a fixed number of self-consistent cycles, the constraint is released and the system is allowed to evolve freely to the closest self-consistent energy minimum. More information about RomeoDFT is provided in Ref.~\citenum{ponet_energy_2024}.
This implementation is made in Quantum ESPRESSO and the process is automatized in \emph{Julia} and can be found on \href{https://github.com/louisponet/RomeoDFTDFT.jl}{GitHub}. 
We use RomeoDFT workflow to determine the ground state at the PBE+$U$ level for 194 magnetic monolayers, identified by Chronos with a filter of up to 12 atoms per primitive cell. 

In order to calculate the $U$ parameter, we use DFPT~\cite{timrov_hubbard_2018, timrov_self-consistent_2021} as implemented in the \textsc{HP} code,~\cite{timrov_hp_2022,cococcioni_linear_2005, leiria_campo_jr_extended_2010} which is part of \textsc{QE}. The Hubbard $U$ parameter is computed in three iterative steps. Within each iteration the $U$ value is updated as the average values of the initial and calculated values in the previous step.~\footnote{Only for special case of CoO$_2$, because the ground state is highly sensitive to the choice of $U$, in order to avoid falling into the wrong states we do not update the $U$ value as the average of the previous step, instead, in each step we use the same calculated value of the $U$ from the previous step.} Our results show that with this approach, the convergence threshold of $U$ for 87$\%$ of the systems is less than 0.5~eV. If the convergence of 0.5~eV has not been reached in three steps, we do another final step. After the calculation of the $U$, the geometry and the Hubbard $U$ are fixed, and we explore the energy landscape to find the global minimum. For the systems with fully occupied $d$ orbitals we use a zero value for the $U$ because the effects of the Hubbard correction at integer occupations are negligible.~\cite{cococcioni_linear_2005}
In the entire process the geometry is fixed to the ground state as identified by Chronos prior to optimization to avoid introducing geometric optimization only on one state. The appropriate approach involves relaxing the geometry across all states within the energy landscape, which can be computationally demanding.\\
To further characterize the magnetic properties of our materials, we mapped their electronic structure onto a model Heisenberg Hamiltonian:
\begin{equation}
H~=~\sum_{ij} J_{ij} \hat S_i\cdot \hat S_j \quad,
\end{equation}
where $J_{ij}$ corresponds to the magnetic exchange coupling between the net spins in sites $i$ and $j$. The magnitude of the spins is included in the $J$.
The sum runs over every spin such that each coupling is counted twice, i.e.  $J_{ij}$ and $J_{ji}$ are both included.
Our convention is such that a negative \( J \) indicates a ferromagnetic coupling, and a positive \( J \) is indicative of antiferromagnetic interaction.
The transition temperature is calculated from mean-field theory through the formula:
\begin{equation}
    T_c~=~\frac{2 \tilde J}{3 k_B} \quad,
\end{equation}
where \( k_B \) is the Boltzmann constant and \( \tilde J~=~\sum_j J_{ij} \).
We calculated the exchange parameters using two approaches. First, we employed the \( |E_{FM} -E_{AFM}| \) method, which maps this total energy difference onto a nearest-neighbor-only model Hamiltonian and the results are shown in~\cref{energy_difference}. Here, the exchange coupling $J$ is assumed to be present between $n$ equivalent nearest neighbors ($\tilde J=nJ$) and given by \( J ~=~ \frac{\Delta E}{2 n N} \), where \( \Delta E \) is the energy difference between self-consistent ferromagnetic and antiferromagnetic spin configurations and $N$ the number of magnetic atoms in the unit cell. 
To account for interactions beyond the first nearest neighbors, we employ the four-state method. In this approach, a large supercell is considered ($4\times4\times1$), and each \( J_{ij} \) exchange interaction is calculated by computing the total energy when the pair of spins in sites \( i \) and \( j \) are in four distinct collinear spin configurations while keeping all other spins ferromagnetically ordered. The four configurations are up-up, down-up, up-down, and down-down concerning the quantization axis. The exchange parameter is then calculated by:~\cite{xiang_predicting_2011}
\begin{equation}
    J_{ij}~=~\frac{1}{8} (E_{\uparrow \uparrow} + E_{\downarrow \downarrow} - E_{\uparrow \downarrow} - E_{\downarrow \uparrow}) 
    \label{4state}
\end{equation}
Finally, to go beyond the mean-field theory and obtain a more precise estimate of the paramagnetic critical temperatures, we use the Monte Carlo Metropolis algorithm as implemented in the \texttt{Vampire package}\,\cite{evans_atomistic_2014, noauthor_vampire_nodate} for a few selected materials. Here, we use periodic boundaries in 2D, an equilibration time of 5.000 Monte Carlo steps, and statistical averaging also over 5.000 steps. The critical temperatures $T_C$ are then obtained by fitting the Monte Carlo magnetization curves to the mean-field formula $M~=~M_0 (1 - T/T_C )^\beta$, where $M_0$ and the critical exponent $\beta$ are also fitting parameters\\

It is important to note the approximations involved in our results. Besides assuming a magnetism driven by localized magnetic moments, the mapping of the \( |E_{FM} - E_{AFM}| \) onto the Heisenberg model considers only nearest neighbors coupling.
This can be regarded as a mean-field approximation. For instance, in MoCl$_3$, the Mo atoms transition from a honeycomb lattice, where each site has three equidistant nearest neighbors, to a triangular lattice of dimers. In this configuration, the Mo–Mo distance within a dimer is 1.06~\AA\ shorter than the distance between dimers (see Supporting Information).

Anisotropy is also neglected in our calculations, and all calculations are done within the non-relativistic limit. While anisotropy could play a significant role in 2D materials, our MC simulations on a model honeycomb lattice considering only the nearest neighbors indicate that $T_c$ is not affected significantly by a typical magnetocrystalline anisotropy. This aligns with recent research, that shows that even for isotropic 2D systems in lab setups short-range interactions can be large enough to stabilize the magnetic order at a finite temperature due to finite effects.~\cite{jenkins_breaking_2022} 

\begin{acknowledgement}
We thank Giovanni Pizzi and Andrea Cepelotti for their crucial roles in helping with scientific discussions and workflow development. We also Guoyuan Liu for fruitful discussions. We acknowledge support from the Swiss National Science Foundation (SNSF), through grant 200021-179138, and its National Centre of Competence in Research (NCCR) MARVEL (grant number 205602). MG acknowledges financial support from Ministero Italiano dell'Università e della Ricerca through the PNRR project "Ecosystem for Sustainable Transition in Emilia-Romagna" (ECS\_00000033-ECOSISTER) and through the PRIN2022 project SECSY (CUP E53D23001700006), both funded by the European Union – NextGenerationEU. Computer time was provided by the Swiss National Supercomputing Centre (CSCS) under project No.~s1073. The authors declare no competing financial interests.
\end{acknowledgement}

\bibliography{My_Library}

\providecommand{\latin}[1]{#1}
\makeatletter
\providecommand{\doi}
  {\begingroup\let\do\@makeother\dospecials
  \catcode`\{=1 \catcode`\}=2 \doi@aux}
\providecommand{\doi@aux}[1]{\endgroup\texttt{#1}}
\makeatother
\providecommand*\mcitethebibliography{\thebibliography}
\csname @ifundefined\endcsname{endmcitethebibliography}  {\let\endmcitethebibliography\endthebibliography}{}
\begin{mcitethebibliography}{114}
\providecommand*\natexlab[1]{#1}
\providecommand*\mciteSetBstSublistMode[1]{}
\providecommand*\mciteSetBstMaxWidthForm[2]{}
\providecommand*\mciteBstWouldAddEndPuncttrue
  {\def\EndOfBibitem{\unskip.}}
\providecommand*\mciteBstWouldAddEndPunctfalse
  {\let\EndOfBibitem\relax}
\providecommand*\mciteSetBstMidEndSepPunct[3]{}
\providecommand*\mciteSetBstSublistLabelBeginEnd[3]{}
\providecommand*\EndOfBibitem{}
\mciteSetBstSublistMode{f}
\mciteSetBstMaxWidthForm{subitem}{(\alph{mcitesubitemcount})}
\mciteSetBstSublistLabelBeginEnd
  {\mcitemaxwidthsubitemform\space}
  {\relax}
  {\relax}

\bibitem[Burch \latin{et~al.}(2018)Burch, Mandrus, and Park]{burch_magnetism_2018}
Burch,~K.~S.; Mandrus,~D.; Park,~J.-G. Magnetism in Two-Dimensional van Der {{Waals}} Materials. \emph{Nature} \textbf{2018}, \emph{563}, 47--52\relax
\mciteBstWouldAddEndPuncttrue
\mciteSetBstMidEndSepPunct{\mcitedefaultmidpunct}
{\mcitedefaultendpunct}{\mcitedefaultseppunct}\relax
\EndOfBibitem
\bibitem[Wang \latin{et~al.}(2022)Wang, {Bedoya-Pinto}, Blei, Dismukes, Hamo, Jenkins, Koperski, Liu, Sun, Telford, Kim, Augustin, Vool, Yin, Li, Falin, Dean, Casanova, Evans, Chshiev, Mishchenko, Petrovic, He, Zhao, Tsen, Gerardot, {Brotons-Gisbert}, Guguchia, Roy, Tongay, Wang, Hasan, Wrachtrup, Yacoby, Fert, Parkin, Novoselov, Dai, Balicas, and Santos]{wang_magnetic_2022}
Wang,~Q.~H.; {Bedoya-Pinto},~A.; Blei,~M.; Dismukes,~A.~H.; Hamo,~A.; Jenkins,~S.; Koperski,~M.; Liu,~Y.; Sun,~Q.-C.; Telford,~E.~J.; Kim,~H.~H.; Augustin,~M.; Vool,~U.; Yin,~J.-X.; Li,~L.~H.; Falin,~A.; Dean,~C.~R.; Casanova,~F.; Evans,~R. F.~L.; Chshiev,~M. \latin{et~al.}  The {{Magnetic Genome}} of {{Two-Dimensional}} van Der {{Waals Materials}}. \emph{ACS Nano} \textbf{2022}, \emph{16}, 6960--7079\relax
\mciteBstWouldAddEndPuncttrue
\mciteSetBstMidEndSepPunct{\mcitedefaultmidpunct}
{\mcitedefaultendpunct}{\mcitedefaultseppunct}\relax
\EndOfBibitem
\bibitem[Rhone \latin{et~al.}(2023)Rhone, Bhattarai, Gavras, Lusch, Salim, Mattheakis, Larson, Krockenberger, and Kaxiras]{rhone_artificial_2023}
Rhone,~T.~D.; Bhattarai,~R.; Gavras,~H.; Lusch,~B.; Salim,~M.; Mattheakis,~M.; Larson,~D.~T.; Krockenberger,~Y.; Kaxiras,~E. Artificial {{Intelligence Guided Studies}} of van Der {{Waals Magnets}}. \emph{Advanced Theory and Simulations} \textbf{2023}, \emph{6}, 2300019\relax
\mciteBstWouldAddEndPuncttrue
\mciteSetBstMidEndSepPunct{\mcitedefaultmidpunct}
{\mcitedefaultendpunct}{\mcitedefaultseppunct}\relax
\EndOfBibitem
\bibitem[Gibertini \latin{et~al.}(2019)Gibertini, Koperski, Morpurgo, and Novoselov]{gibertini_magnetic_2019}
Gibertini,~M.; Koperski,~M.; Morpurgo,~A.~F.; Novoselov,~K.~S. Magnetic {{2D}} Materials and Heterostructures. \emph{Nature Nanotechnology} \textbf{2019}, \emph{14}, 408--419\relax
\mciteBstWouldAddEndPuncttrue
\mciteSetBstMidEndSepPunct{\mcitedefaultmidpunct}
{\mcitedefaultendpunct}{\mcitedefaultseppunct}\relax
\EndOfBibitem
\bibitem[S{\o}dequist and Olsen(2023)S{\o}dequist, and Olsen]{sodequist_type_2023}
S{\o}dequist,~J.; Olsen,~T. Type {{II}} Multiferroic Order in Two-Dimensional Transition Metal Halides from First Principles Spin-Spiral Calculations. 2023\relax
\mciteBstWouldAddEndPuncttrue
\mciteSetBstMidEndSepPunct{\mcitedefaultmidpunct}
{\mcitedefaultendpunct}{\mcitedefaultseppunct}\relax
\EndOfBibitem
\bibitem[Ovesen and Olsen(2024)Ovesen, and Olsen]{ovesen_orbital_2024}
Ovesen,~M.; Olsen,~T. Orbital Magnetization in Two-Dimensional Materials from High-Throughput Computational Screening. \emph{2D Materials} \textbf{2024}, \emph{11}, 045010\relax
\mciteBstWouldAddEndPuncttrue
\mciteSetBstMidEndSepPunct{\mcitedefaultmidpunct}
{\mcitedefaultendpunct}{\mcitedefaultseppunct}\relax
\EndOfBibitem
\bibitem[Xin \latin{et~al.}(2023)Xin, Yin, Song, Fan, Song, and Pan]{xin_machine_2023}
Xin,~C.; Yin,~Y.; Song,~B.; Fan,~Z.; Song,~Y.; Pan,~F. Machine Learning-Accelerated Discovery of Novel {{2D}} Ferromagnetic Materials with Strong Magnetization. \emph{Chip} \textbf{2023}, \emph{2}, 100071\relax
\mciteBstWouldAddEndPuncttrue
\mciteSetBstMidEndSepPunct{\mcitedefaultmidpunct}
{\mcitedefaultendpunct}{\mcitedefaultseppunct}\relax
\EndOfBibitem
\bibitem[Shen \latin{et~al.}(2022)Shen, Su, and He]{shen_high-throughput_2022}
Shen,~Z.-X.; Su,~C.; He,~L. High-Throughput Computation and Structure Prototype Analysis for Two-Dimensional Ferromagnetic Materials. \emph{npj Computational Materials} \textbf{2022}, \emph{8}, 132\relax
\mciteBstWouldAddEndPuncttrue
\mciteSetBstMidEndSepPunct{\mcitedefaultmidpunct}
{\mcitedefaultendpunct}{\mcitedefaultseppunct}\relax
\EndOfBibitem
\bibitem[Xia \latin{et~al.}(2022)Xia, Sakurai, Balasubramanian, Liao, Wang, Zhang, Sun, Ho, Chelikowsky, Sellmyer, and Wang]{xia_accelerating_2022}
Xia,~W.; Sakurai,~M.; Balasubramanian,~B.; Liao,~T.; Wang,~R.; Zhang,~C.; Sun,~H.; Ho,~K.-M.; Chelikowsky,~J.~R.; Sellmyer,~D.~J.; Wang,~C.-Z. Accelerating the Discovery of Novel Magnetic Materials Using Machine Learning--Guided Adaptive Feedback. \emph{Proceedings of the National Academy of Sciences} \textbf{2022}, \emph{119}, e2204485119\relax
\mciteBstWouldAddEndPuncttrue
\mciteSetBstMidEndSepPunct{\mcitedefaultmidpunct}
{\mcitedefaultendpunct}{\mcitedefaultseppunct}\relax
\EndOfBibitem
\bibitem[Huang \latin{et~al.}(2017)Huang, Clark, {Navarro-Moratalla}, Klein, Cheng, Seyler, Zhong, Schmidgall, McGuire, Cobden, Yao, Xiao, {Jarillo-Herrero}, and Xu]{huang_layer-dependent_2017}
Huang,~B.; Clark,~G.; {Navarro-Moratalla},~E.; Klein,~D.~R.; Cheng,~R.; Seyler,~K.~L.; Zhong,~D.; Schmidgall,~E.; McGuire,~M.~A.; Cobden,~D.~H.; Yao,~W.; Xiao,~D.; {Jarillo-Herrero},~P.; Xu,~X. Layer-Dependent Ferromagnetism in a van Der {{Waals}} Crystal down to the Monolayer Limit. \emph{Nature} \textbf{2017}, \emph{546}, 270--273\relax
\mciteBstWouldAddEndPuncttrue
\mciteSetBstMidEndSepPunct{\mcitedefaultmidpunct}
{\mcitedefaultendpunct}{\mcitedefaultseppunct}\relax
\EndOfBibitem
\bibitem[{Bedoya-Pinto} \latin{et~al.}(2021){Bedoya-Pinto}, Ji, Pandeya, Gargiani, Valvidares, Sessi, Taylor, Radu, Chang, and Parkin]{bedoya-pinto_intrinsic_2021}
{Bedoya-Pinto},~A.; Ji,~J.-R.; Pandeya,~A.~K.; Gargiani,~P.; Valvidares,~M.; Sessi,~P.; Taylor,~J.~M.; Radu,~F.; Chang,~K.; Parkin,~S. S.~P. Intrinsic {{2D-XY}} Ferromagnetism in a van Der {{Waals}} Monolayer. \emph{Science} \textbf{2021}, \emph{374}, 616--620\relax
\mciteBstWouldAddEndPuncttrue
\mciteSetBstMidEndSepPunct{\mcitedefaultmidpunct}
{\mcitedefaultendpunct}{\mcitedefaultseppunct}\relax
\EndOfBibitem
\bibitem[Posey \latin{et~al.}(2024)Posey, Turkel, Rezaee, Devarakonda, Kundu, Ong, Thinel, Chica, Vitalone, Jing, Xu, Needell, Meirzadeh, Feuer, Jindal, Cui, Valla, Thunstr{\"o}m, Yilmaz, Vescovo, Graf, Zhu, Scheie, May, Eriksson, Basov, Dean, Rubio, Kim, Ziebel, Millis, Pasupathy, and Roy]{posey_two-dimensional_2024}
Posey,~V.~A.; Turkel,~S.; Rezaee,~M.; Devarakonda,~A.; Kundu,~A.~K.; Ong,~C.~S.; Thinel,~M.; Chica,~D.~G.; Vitalone,~R.~A.; Jing,~R.; Xu,~S.; Needell,~D.~R.; Meirzadeh,~E.; Feuer,~M.~L.; Jindal,~A.; Cui,~X.; Valla,~T.; Thunstr{\"o}m,~P.; Yilmaz,~T.; Vescovo,~E. \latin{et~al.}  Two-Dimensional Heavy Fermions in the van Der {{Waals}} Metal {{CeSiI}}. \emph{Nature} \textbf{2024}, \emph{625}, 483--488\relax
\mciteBstWouldAddEndPuncttrue
\mciteSetBstMidEndSepPunct{\mcitedefaultmidpunct}
{\mcitedefaultendpunct}{\mcitedefaultseppunct}\relax
\EndOfBibitem
\bibitem[Mounet \latin{et~al.}(2018)Mounet, Gibertini, Schwaller, Campi, Merkys, Marrazzo, Sohier, Castelli, Cepellotti, Pizzi, and Marzari]{mounet_two-dimensional_2018}
Mounet,~N.; Gibertini,~M.; Schwaller,~P.; Campi,~D.; Merkys,~A.; Marrazzo,~A.; Sohier,~T.; Castelli,~I.~E.; Cepellotti,~A.; Pizzi,~G.; Marzari,~N. Two-Dimensional Materials from High-Throughput Computational Exfoliation of Experimentally Known Compounds. \emph{Nature Nanotechnology} \textbf{2018}, \emph{13}, 246--252\relax
\mciteBstWouldAddEndPuncttrue
\mciteSetBstMidEndSepPunct{\mcitedefaultmidpunct}
{\mcitedefaultendpunct}{\mcitedefaultseppunct}\relax
\EndOfBibitem
\bibitem[Campi \latin{et~al.}(2023)Campi, Mounet, Gibertini, Pizzi, and Marzari]{campi_expansion_2023}
Campi,~D.; Mounet,~N.; Gibertini,~M.; Pizzi,~G.; Marzari,~N. Expansion of the {{Materials Cloud 2D Database}}. \emph{ACS Nano} \textbf{2023}, \emph{17}, 11268--11278\relax
\mciteBstWouldAddEndPuncttrue
\mciteSetBstMidEndSepPunct{\mcitedefaultmidpunct}
{\mcitedefaultendpunct}{\mcitedefaultseppunct}\relax
\EndOfBibitem
\bibitem[Leb{\`e}gue \latin{et~al.}(2013)Leb{\`e}gue, Bj{\"o}rkman, Klintenberg, Nieminen, and Eriksson]{lebegue_two-dimensional_2013}
Leb{\`e}gue,~S.; Bj{\"o}rkman,~T.; Klintenberg,~M.; Nieminen,~R.~M.; Eriksson,~O. Two-{{Dimensional Materials}} from {{Data Filtering}} and {{{\emph{Ab Initio}}}} {{Calculations}}. \emph{Physical Review X} \textbf{2013}, \emph{3}, 031002\relax
\mciteBstWouldAddEndPuncttrue
\mciteSetBstMidEndSepPunct{\mcitedefaultmidpunct}
{\mcitedefaultendpunct}{\mcitedefaultseppunct}\relax
\EndOfBibitem
\bibitem[Rasmussen and Thygesen(2015)Rasmussen, and Thygesen]{rasmussen_computational_2015}
Rasmussen,~F.~A.; Thygesen,~K.~S. Computational {{2D Materials Database}}: {{Electronic Structure}} of {{Transition-Metal Dichalcogenides}} and {{Oxides}}. \emph{The Journal of Physical Chemistry C} \textbf{2015}, \emph{119}, 13169--13183\relax
\mciteBstWouldAddEndPuncttrue
\mciteSetBstMidEndSepPunct{\mcitedefaultmidpunct}
{\mcitedefaultendpunct}{\mcitedefaultseppunct}\relax
\EndOfBibitem
\bibitem[Choudhary \latin{et~al.}(2017)Choudhary, Kalish, Beams, and Tavazza]{choudhary_high-throughput_2017}
Choudhary,~K.; Kalish,~I.; Beams,~R.; Tavazza,~F. High-Throughput {{Identification}} and {{Characterization}} of {{Two-dimensional Materials}} Using {{Density}} Functional Theory. \emph{Scientific Reports} \textbf{2017}, \emph{7}, 5179\relax
\mciteBstWouldAddEndPuncttrue
\mciteSetBstMidEndSepPunct{\mcitedefaultmidpunct}
{\mcitedefaultendpunct}{\mcitedefaultseppunct}\relax
\EndOfBibitem
\bibitem[Ashton \latin{et~al.}(2017)Ashton, Paul, Sinnott, and Hennig]{ashton_topology-scaling_2017}
Ashton,~M.; Paul,~J.; Sinnott,~S.~B.; Hennig,~R.~G. Topology-{{Scaling Identification}} of {{Layered Solids}} and {{Stable Exfoliated 2D Materials}}. \emph{Physical Review Letters} \textbf{2017}, \emph{118}, 106101\relax
\mciteBstWouldAddEndPuncttrue
\mciteSetBstMidEndSepPunct{\mcitedefaultmidpunct}
{\mcitedefaultendpunct}{\mcitedefaultseppunct}\relax
\EndOfBibitem
\bibitem[Cheon \latin{et~al.}(2017)Cheon, Duerloo, Sendek, Porter, Chen, and Reed]{cheon_data_2017}
Cheon,~G.; Duerloo,~K.-A.~N.; Sendek,~A.~D.; Porter,~C.; Chen,~Y.; Reed,~E.~J. Data {{Mining}} for {{New Two-}} and {{One-Dimensional Weakly Bonded Solids}} and {{Lattice-Commensurate Heterostructures}}. \emph{Nano Letters} \textbf{2017}, \emph{17}, 1915--1923\relax
\mciteBstWouldAddEndPuncttrue
\mciteSetBstMidEndSepPunct{\mcitedefaultmidpunct}
{\mcitedefaultendpunct}{\mcitedefaultseppunct}\relax
\EndOfBibitem
\bibitem[Zhou \latin{et~al.}(2019)Zhou, Shen, Costa, Persson, Ong, Huck, Lu, Ma, Chen, Tang, and Feng]{zhou_2dmatpedia_2019}
Zhou,~J.; Shen,~L.; Costa,~M.~D.; Persson,~K.~A.; Ong,~S.~P.; Huck,~P.; Lu,~Y.; Ma,~X.; Chen,~Y.; Tang,~H.; Feng,~Y.~P. {{2DMatPedia}}, an Open Computational Database of Two-Dimensional Materials from Top-down and Bottom-up Approaches. \emph{Scientific Data} \textbf{2019}, \emph{6}, 86\relax
\mciteBstWouldAddEndPuncttrue
\mciteSetBstMidEndSepPunct{\mcitedefaultmidpunct}
{\mcitedefaultendpunct}{\mcitedefaultseppunct}\relax
\EndOfBibitem
\bibitem[Haastrup \latin{et~al.}(2018)Haastrup, Strange, Pandey, Deilmann, Schmidt, Hinsche, Gjerding, Torelli, Larsen, {Riis-Jensen}, Gath, Jacobsen, J{\o}rgen~Mortensen, Olsen, and Thygesen]{haastrup_computational_2018}
Haastrup,~S.; Strange,~M.; Pandey,~M.; Deilmann,~T.; Schmidt,~P.~S.; Hinsche,~N.~F.; Gjerding,~M.~N.; Torelli,~D.; Larsen,~P.~M.; {Riis-Jensen},~A.~C.; Gath,~J.; Jacobsen,~K.~W.; J{\o}rgen~Mortensen,~J.; Olsen,~T.; Thygesen,~K.~S. The {{Computational 2D Materials Database}}: High-Throughput Modeling and Discovery of Atomically Thin Crystals. \emph{2D Materials} \textbf{2018}, \emph{5}, 042002\relax
\mciteBstWouldAddEndPuncttrue
\mciteSetBstMidEndSepPunct{\mcitedefaultmidpunct}
{\mcitedefaultendpunct}{\mcitedefaultseppunct}\relax
\EndOfBibitem
\bibitem[Gjerding \latin{et~al.}(2021)Gjerding, Taghizadeh, Rasmussen, Ali, Bertoldo, Deilmann, Kn{\o}sgaard, Kruse, Larsen, Manti, Pedersen, Petralanda, Skovhus, Svendsen, Mortensen, Olsen, and Thygesen]{gjerding_recent_2021}
Gjerding,~M.~N.; Taghizadeh,~A.; Rasmussen,~A.; Ali,~S.; Bertoldo,~F.; Deilmann,~T.; Kn{\o}sgaard,~N.~R.; Kruse,~M.; Larsen,~A.~H.; Manti,~S.; Pedersen,~T.~G.; Petralanda,~U.; Skovhus,~T.; Svendsen,~M.~K.; Mortensen,~J.~J.; Olsen,~T.; Thygesen,~K.~S. Recent Progress of the {{Computational 2D Materials Database}} ({{C2DB}}). \emph{2D Materials} \textbf{2021}, \emph{8}, 044002\relax
\mciteBstWouldAddEndPuncttrue
\mciteSetBstMidEndSepPunct{\mcitedefaultmidpunct}
{\mcitedefaultendpunct}{\mcitedefaultseppunct}\relax
\EndOfBibitem
\bibitem[Torelli \latin{et~al.}(2019)Torelli, Thygesen, and Olsen]{torelli_high_2019}
Torelli,~D.; Thygesen,~K.~S.; Olsen,~T. High Throughput Computational Screening for {{2D}} Ferromagnetic Materials: The Critical Role of Anisotropy and Local Correlations. \emph{2D Materials} \textbf{2019}, \emph{6}, 045018\relax
\mciteBstWouldAddEndPuncttrue
\mciteSetBstMidEndSepPunct{\mcitedefaultmidpunct}
{\mcitedefaultendpunct}{\mcitedefaultseppunct}\relax
\EndOfBibitem
\bibitem[Kabiraj \latin{et~al.}(2020)Kabiraj, Kumar, and Mahapatra]{kabiraj_high-throughput_2020}
Kabiraj,~A.; Kumar,~M.; Mahapatra,~S. High-Throughput Discovery of High {{Curie}} Point Two-Dimensional Ferromagnetic Materials. \emph{npj Computational Materials} \textbf{2020}, \emph{6}, 35\relax
\mciteBstWouldAddEndPuncttrue
\mciteSetBstMidEndSepPunct{\mcitedefaultmidpunct}
{\mcitedefaultendpunct}{\mcitedefaultseppunct}\relax
\EndOfBibitem
\bibitem[Torelli and Olsen(2020)Torelli, and Olsen]{torelli_first_2020}
Torelli,~D.; Olsen,~T. First Principles {{Heisenberg}} Models of {{2D}} Magnetic Materials: The Importance of Quantum Corrections to the Exchange Coupling. \emph{Journal of Physics: Condensed Matter} \textbf{2020}, \emph{32}, 335802\relax
\mciteBstWouldAddEndPuncttrue
\mciteSetBstMidEndSepPunct{\mcitedefaultmidpunct}
{\mcitedefaultendpunct}{\mcitedefaultseppunct}\relax
\EndOfBibitem
\bibitem[Torelli and Olsen(2018)Torelli, and Olsen]{torelli_calculating_2018}
Torelli,~D.; Olsen,~T. Calculating Critical Temperatures for Ferromagnetic Order in Two-Dimensional Materials. \emph{2D Materials} \textbf{2018}, \emph{6}, 015028\relax
\mciteBstWouldAddEndPuncttrue
\mciteSetBstMidEndSepPunct{\mcitedefaultmidpunct}
{\mcitedefaultendpunct}{\mcitedefaultseppunct}\relax
\EndOfBibitem
\bibitem[Kulik \latin{et~al.}(2006)Kulik, Cococcioni, Scherlis, and Marzari]{kulik_density_2006}
Kulik,~H.~J.; Cococcioni,~M.; Scherlis,~D.~A.; Marzari,~N. Density {{Functional Theory}} in {{Transition-Metal Chemistry}}: {{A Self-Consistent Hubbard U Approach}}. \emph{Physical Review Letters} \textbf{2006}, \emph{97}, 103001\relax
\mciteBstWouldAddEndPuncttrue
\mciteSetBstMidEndSepPunct{\mcitedefaultmidpunct}
{\mcitedefaultendpunct}{\mcitedefaultseppunct}\relax
\EndOfBibitem
\bibitem[Anisimov \latin{et~al.}(1991)Anisimov, Zaanen, and Andersen]{anisimov_band_1991}
Anisimov,~V.~I.; Zaanen,~J.; Andersen,~O.~K. Band Theory and {{Mott}} Insulators: {{Hubbard}} {{{\emph{U}}}} Instead of {{Stoner}} {{{\emph{I}}}}. \emph{Physical Review B} \textbf{1991}, \emph{44}, 943--954\relax
\mciteBstWouldAddEndPuncttrue
\mciteSetBstMidEndSepPunct{\mcitedefaultmidpunct}
{\mcitedefaultendpunct}{\mcitedefaultseppunct}\relax
\EndOfBibitem
\bibitem[Anisimov \latin{et~al.}(1997)Anisimov, Aryasetiawan, and Lichtenstein]{anisimov_first-principles_1997}
Anisimov,~V.~I.; Aryasetiawan,~F.; Lichtenstein,~A.~I. First-Principles Calculations of the Electronic Structure and Spectra of Strongly Correlated Systems: The {{{\textbf{LDA}}}} + {{{\emph{U}}}} Method. \emph{Journal of Physics: Condensed Matter} \textbf{1997}, \emph{9}, 767--808\relax
\mciteBstWouldAddEndPuncttrue
\mciteSetBstMidEndSepPunct{\mcitedefaultmidpunct}
{\mcitedefaultendpunct}{\mcitedefaultseppunct}\relax
\EndOfBibitem
\bibitem[Dudarev \latin{et~al.}(1998)Dudarev, Botton, Savrasov, Humphreys, and Sutton]{dudarev_electron-energy-loss_1998}
Dudarev,~S.~L.; Botton,~G.~A.; Savrasov,~S.~Y.; Humphreys,~C.~J.; Sutton,~A.~P. Electron-Energy-Loss Spectra and the Structural Stability of Nickel Oxide: {{An LSDA}}+{{U}} Study. \emph{Physical Review B} \textbf{1998}, \emph{57}, 1505--1509\relax
\mciteBstWouldAddEndPuncttrue
\mciteSetBstMidEndSepPunct{\mcitedefaultmidpunct}
{\mcitedefaultendpunct}{\mcitedefaultseppunct}\relax
\EndOfBibitem
\bibitem[Cococcioni and {de Gironcoli}(2005)Cococcioni, and {de Gironcoli}]{cococcioni_linear_2005}
Cococcioni,~M.; {de Gironcoli},~S. Linear Response Approach to the Calculation of the Effective Interaction Parameters in the {{LDA}}+{{U}} Method. \emph{Physical Review B} \textbf{2005}, \emph{71}, 035105\relax
\mciteBstWouldAddEndPuncttrue
\mciteSetBstMidEndSepPunct{\mcitedefaultmidpunct}
{\mcitedefaultendpunct}{\mcitedefaultseppunct}\relax
\EndOfBibitem
\bibitem[Moore \latin{et~al.}(2022)Moore, Horton, Ganose, Siron, Linscott, O'Regan, and Persson]{moore_high-throughput_2022}
Moore,~G.~C.; Horton,~M.~K.; Ganose,~A.~M.; Siron,~M.; Linscott,~E.; O'Regan,~D.~D.; Persson,~K.~A. High-Throughput Determination of {{Hubbard U}} and {{Hund J}} Values for Transition Metal Oxides via Linear Response Formalism. 2022\relax
\mciteBstWouldAddEndPuncttrue
\mciteSetBstMidEndSepPunct{\mcitedefaultmidpunct}
{\mcitedefaultendpunct}{\mcitedefaultseppunct}\relax
\EndOfBibitem
\bibitem[Timrov \latin{et~al.}(2021)Timrov, Marzari, and Cococcioni]{timrov_self-consistent_2021}
Timrov,~I.; Marzari,~N.; Cococcioni,~M. Self-Consistent {{Hubbard}} Parameters from Density-Functional Perturbation Theory in the Ultrasoft and Projector-Augmented Wave Formulations. \emph{Physical Review B} \textbf{2021}, \emph{103}, 045141\relax
\mciteBstWouldAddEndPuncttrue
\mciteSetBstMidEndSepPunct{\mcitedefaultmidpunct}
{\mcitedefaultendpunct}{\mcitedefaultseppunct}\relax
\EndOfBibitem
\bibitem[Timrov \latin{et~al.}(2022)Timrov, Marzari, and Cococcioni]{timrov_hp_2022}
Timrov,~I.; Marzari,~N.; Cococcioni,~M. {{HP}} -- {{A}} Code for the Calculation of {{Hubbard}} Parameters Using Density-Functional Perturbation Theory. \emph{Computer Physics Communications} \textbf{2022}, \emph{279}, 108455\relax
\mciteBstWouldAddEndPuncttrue
\mciteSetBstMidEndSepPunct{\mcitedefaultmidpunct}
{\mcitedefaultendpunct}{\mcitedefaultseppunct}\relax
\EndOfBibitem
\bibitem[Timrov \latin{et~al.}(2018)Timrov, Marzari, and Cococcioni]{timrov_hubbard_2018}
Timrov,~I.; Marzari,~N.; Cococcioni,~M. Hubbard Parameters from Density-Functional Perturbation Theory. \emph{Physical Review B} \textbf{2018}, \emph{98}, 085127\relax
\mciteBstWouldAddEndPuncttrue
\mciteSetBstMidEndSepPunct{\mcitedefaultmidpunct}
{\mcitedefaultendpunct}{\mcitedefaultseppunct}\relax
\EndOfBibitem
\bibitem[bas(2025)]{bastonero_first-principles_2025}
https://archive.materialscloud.org/records/4x37g-nre30. 2025\relax
\mciteBstWouldAddEndPuncttrue
\mciteSetBstMidEndSepPunct{\mcitedefaultmidpunct}
{\mcitedefaultendpunct}{\mcitedefaultseppunct}\relax
\EndOfBibitem
\bibitem[Leiria Campo~Jr and Cococcioni(2010)Leiria Campo~Jr, and Cococcioni]{leiria_campo_jr_extended_2010}
Leiria Campo~Jr,~V.; Cococcioni,~M. Extended {{DFT}} + {{{\emph{U}}}} + {{{\emph{V}}}} Method with On-Site and Inter-Site Electronic Interactions. \emph{Journal of Physics: Condensed Matter} \textbf{2010}, \emph{22}, 055602\relax
\mciteBstWouldAddEndPuncttrue
\mciteSetBstMidEndSepPunct{\mcitedefaultmidpunct}
{\mcitedefaultendpunct}{\mcitedefaultseppunct}\relax
\EndOfBibitem
\bibitem[Haddadi \latin{et~al.}(2024)Haddadi, Linscott, Timrov, Marzari, and Gibertini]{haddadi_-site_2024}
Haddadi,~F.; Linscott,~E.; Timrov,~I.; Marzari,~N.; Gibertini,~M. On-Site and Intersite {{Hubbard}} Corrections in Magnetic Monolayers: {{The}} Case of {{FePS}} 3 and {{CrI}} 3. \emph{Physical Review Materials} \textbf{2024}, \emph{8}, 014007\relax
\mciteBstWouldAddEndPuncttrue
\mciteSetBstMidEndSepPunct{\mcitedefaultmidpunct}
{\mcitedefaultendpunct}{\mcitedefaultseppunct}\relax
\EndOfBibitem
\bibitem[Meredig \latin{et~al.}(2010)Meredig, Thompson, Hansen, Wolverton, and {van de Walle}]{meredig_method_2010}
Meredig,~B.; Thompson,~A.; Hansen,~H.~A.; Wolverton,~C.; {van de Walle},~A. Method for Locating Low-Energy Solutions within {{DFT}} + {{U}}. \emph{Physical Review B} \textbf{2010}, \emph{82}, 195128\relax
\mciteBstWouldAddEndPuncttrue
\mciteSetBstMidEndSepPunct{\mcitedefaultmidpunct}
{\mcitedefaultendpunct}{\mcitedefaultseppunct}\relax
\EndOfBibitem
\bibitem[Zhou and Ozoli{\c n}{\v s}(2009)Zhou, and Ozoli{\c n}{\v s}]{zhou_obtaining_2009}
Zhou,~F.; Ozoli{\c n}{\v s},~V. Obtaining Correct Orbital Ground States in f -Electron Systems Using a Nonspherical Self-Interaction-Corrected {{LDA}} + {{U}} Method. \emph{Physical Review B} \textbf{2009}, \emph{80}, 125127\relax
\mciteBstWouldAddEndPuncttrue
\mciteSetBstMidEndSepPunct{\mcitedefaultmidpunct}
{\mcitedefaultendpunct}{\mcitedefaultseppunct}\relax
\EndOfBibitem
\bibitem[Ylvisaker \latin{et~al.}(2009)Ylvisaker, Pickett, and Koepernik]{ylvisaker_anisotropy_2009}
Ylvisaker,~E.~R.; Pickett,~W.~E.; Koepernik,~K. Anisotropy and Magnetism in the {{LSDA}} + {{U}} Method. \emph{Physical Review B} \textbf{2009}, \emph{79}, 035103\relax
\mciteBstWouldAddEndPuncttrue
\mciteSetBstMidEndSepPunct{\mcitedefaultmidpunct}
{\mcitedefaultendpunct}{\mcitedefaultseppunct}\relax
\EndOfBibitem
\bibitem[Jollet \latin{et~al.}(2009)Jollet, Jomard, Amadon, Crocombette, and Torumba]{jollet_hybrid_2009}
Jollet,~F.; Jomard,~G.; Amadon,~B.; Crocombette,~J.~P.; Torumba,~D. Hybrid Functional for Correlated Electrons in the Projector Augmented-Wave Formalism: {{Study}} of Multiple Minima for Actinide Oxides. \emph{Physical Review B} \textbf{2009}, \emph{80}, 235109\relax
\mciteBstWouldAddEndPuncttrue
\mciteSetBstMidEndSepPunct{\mcitedefaultmidpunct}
{\mcitedefaultendpunct}{\mcitedefaultseppunct}\relax
\EndOfBibitem
\bibitem[Jomard \latin{et~al.}(2008)Jomard, Amadon, Bottin, and Torrent]{jomard_structural_2008}
Jomard,~G.; Amadon,~B.; Bottin,~F.; Torrent,~M. Structural, Thermodynamic, and Electronic Properties of Plutonium Oxides from First Principles. \emph{Physical Review B} \textbf{2008}, \emph{78}, 075125\relax
\mciteBstWouldAddEndPuncttrue
\mciteSetBstMidEndSepPunct{\mcitedefaultmidpunct}
{\mcitedefaultendpunct}{\mcitedefaultseppunct}\relax
\EndOfBibitem
\bibitem[Amadon \latin{et~al.}(2008)Amadon, Jollet, and Torrent]{amadon__2008}
Amadon,~B.; Jollet,~F.; Torrent,~M. {$\gamma$} and {$\beta$} Cerium: {{LDA}} + {{U}} Calculations of Ground-State Parameters. \emph{Physical Review B} \textbf{2008}, \emph{77}, 155104\relax
\mciteBstWouldAddEndPuncttrue
\mciteSetBstMidEndSepPunct{\mcitedefaultmidpunct}
{\mcitedefaultendpunct}{\mcitedefaultseppunct}\relax
\EndOfBibitem
\bibitem[Zhang \latin{et~al.}(2009)Zhang, Koepernik, Richter, and Eschrig]{zhang_magnetic_2009}
Zhang,~W.; Koepernik,~K.; Richter,~M.; Eschrig,~H. Magnetic Phase Transition in {{CoO}} under High Pressure: {{A}} Challenge for {{LSDA}} + {{U}}. \emph{Physical Review B} \textbf{2009}, \emph{79}, 155123\relax
\mciteBstWouldAddEndPuncttrue
\mciteSetBstMidEndSepPunct{\mcitedefaultmidpunct}
{\mcitedefaultendpunct}{\mcitedefaultseppunct}\relax
\EndOfBibitem
\bibitem[Kasinathan \latin{et~al.}(2007)Kasinathan, Koepernik, Nitzsche, and Rosner]{kasinathan_ferromagnetism_2007}
Kasinathan,~D.; Koepernik,~K.; Nitzsche,~U.; Rosner,~H. Ferromagnetism {{Induced}} by {{Orbital Order}} in the {{Charge-Transfer Insulator Cs}} 2 {{AgF}} 4 : {{An Electronic Structure Study}}. \emph{Physical Review Letters} \textbf{2007}, \emph{99}, 247210\relax
\mciteBstWouldAddEndPuncttrue
\mciteSetBstMidEndSepPunct{\mcitedefaultmidpunct}
{\mcitedefaultendpunct}{\mcitedefaultseppunct}\relax
\EndOfBibitem
\bibitem[Shick \latin{et~al.}(2004)Shick, Jani{\v s}, Drchal, and Pickett]{shick_spin_2004}
Shick,~A.~B.; Jani{\v s},~V.; Drchal,~V.; Pickett,~W.~E. Spin and Orbital Magnetic State of {{U Ge}} 2 under Pressure. \emph{Physical Review B} \textbf{2004}, \emph{70}, 134506\relax
\mciteBstWouldAddEndPuncttrue
\mciteSetBstMidEndSepPunct{\mcitedefaultmidpunct}
{\mcitedefaultendpunct}{\mcitedefaultseppunct}\relax
\EndOfBibitem
\bibitem[Ponet \latin{et~al.}(2024)Ponet, Di~Lucente, and Marzari]{ponet_energy_2024}
Ponet,~L.; Di~Lucente,~E.; Marzari,~N. The Energy Landscape of Magnetic Materials. \emph{npj Computational Materials} \textbf{2024}, \emph{10}, 151\relax
\mciteBstWouldAddEndPuncttrue
\mciteSetBstMidEndSepPunct{\mcitedefaultmidpunct}
{\mcitedefaultendpunct}{\mcitedefaultseppunct}\relax
\EndOfBibitem
\bibitem[Horton \latin{et~al.}(2019)Horton, Montoya, Liu, and Persson]{horton_high-throughput_2019}
Horton,~M.~K.; Montoya,~J.~H.; Liu,~M.; Persson,~K.~A. High-Throughput Prediction of the Ground-State Collinear Magnetic Order of Inorganic Materials Using {{Density Functional Theory}}. \emph{npj Computational Materials} \textbf{2019}, \emph{5}, 64\relax
\mciteBstWouldAddEndPuncttrue
\mciteSetBstMidEndSepPunct{\mcitedefaultmidpunct}
{\mcitedefaultendpunct}{\mcitedefaultseppunct}\relax
\EndOfBibitem
\bibitem[Payne \latin{et~al.}(2019)Payne, {Aveda{\~n}o-Franco}, He, Bousquet, and Romero]{payne_optimizing_2019}
Payne,~A.; {Aveda{\~n}o-Franco},~G.; He,~X.; Bousquet,~E.; Romero,~A.~H. Optimizing the Orbital Occupation in the Multiple Minima Problem of Magnetic Materials from the Metaheuristic Firefly Algorithm. \emph{Physical Chemistry Chemical Physics} \textbf{2019}, \emph{21}, 21932--21941\relax
\mciteBstWouldAddEndPuncttrue
\mciteSetBstMidEndSepPunct{\mcitedefaultmidpunct}
{\mcitedefaultendpunct}{\mcitedefaultseppunct}\relax
\EndOfBibitem
\bibitem[Allen and Watson(2014)Allen, and Watson]{allen_occupation_2014}
Allen,~J.~P.; Watson,~G.~W. Occupation Matrix Control of d- and f-Electron Localisations Using {{DFT}}+{{U}}. \emph{Phys. Chem. Chem. Phys.} \textbf{2014}, \emph{16}, 21016--21031\relax
\mciteBstWouldAddEndPuncttrue
\mciteSetBstMidEndSepPunct{\mcitedefaultmidpunct}
{\mcitedefaultendpunct}{\mcitedefaultseppunct}\relax
\EndOfBibitem
\bibitem[Dorado \latin{et~al.}(2010)Dorado, Jomard, Freyss, and Bertolus]{dorado_stability_2010}
Dorado,~B.; Jomard,~G.; Freyss,~M.; Bertolus,~M. Stability of Oxygen Point Defects in {{UO}} 2 by First-Principles {{DFT}} + {{U}} Calculations: {{Occupation}} Matrix Control and {{Jahn-Teller}} Distortion. \emph{Physical Review B} \textbf{2010}, \emph{82}, 035114\relax
\mciteBstWouldAddEndPuncttrue
\mciteSetBstMidEndSepPunct{\mcitedefaultmidpunct}
{\mcitedefaultendpunct}{\mcitedefaultseppunct}\relax
\EndOfBibitem
\bibitem[Dorado \latin{et~al.}(2009)Dorado, Amadon, Freyss, and Bertolus]{dorado_dft_2009}
Dorado,~B.; Amadon,~B.; Freyss,~M.; Bertolus,~M. {{DFT}} + {{U}} Calculations of the Ground State and Metastable States of Uranium Dioxide. \emph{Physical Review B} \textbf{2009}, \emph{79}, 235125\relax
\mciteBstWouldAddEndPuncttrue
\mciteSetBstMidEndSepPunct{\mcitedefaultmidpunct}
{\mcitedefaultendpunct}{\mcitedefaultseppunct}\relax
\EndOfBibitem
\bibitem[{Tellez-Mora} \latin{et~al.}(2024){Tellez-Mora}, He, Bousquet, Wirtz, and Romero]{tellez-mora_systematic_2024}
{Tellez-Mora},~A.; He,~X.; Bousquet,~E.; Wirtz,~L.; Romero,~A.~H. Systematic Determination of a Material's Magnetic Ground State from First Principles. \emph{npj Computational Materials} \textbf{2024}, \emph{10}, 20\relax
\mciteBstWouldAddEndPuncttrue
\mciteSetBstMidEndSepPunct{\mcitedefaultmidpunct}
{\mcitedefaultendpunct}{\mcitedefaultseppunct}\relax
\EndOfBibitem
\bibitem[S{\o}dequist and Olsen(2024)S{\o}dequist, and Olsen]{sodequist_magnetic_2024}
S{\o}dequist,~J.; Olsen,~T. Magnetic Order in the Computational {{2D}} Materials Database ({{C2DB}}) from High Throughput Spin Spiral Calculations. \emph{npj Computational Materials} \textbf{2024}, \emph{10}, 170\relax
\mciteBstWouldAddEndPuncttrue
\mciteSetBstMidEndSepPunct{\mcitedefaultmidpunct}
{\mcitedefaultendpunct}{\mcitedefaultseppunct}\relax
\EndOfBibitem
\bibitem[Huebsch \latin{et~al.}(2021)Huebsch, Nomoto, Suzuki, and Arita]{huebsch_benchmark_2021}
Huebsch,~M.-T.; Nomoto,~T.; Suzuki,~M.-T.; Arita,~R. Benchmark for {{{\emph{Ab Initio}}}} {{Prediction}} of {{Magnetic Structures Based}} on {{Cluster-Multipole Theory}}. \emph{Physical Review X} \textbf{2021}, \emph{11}, 011031\relax
\mciteBstWouldAddEndPuncttrue
\mciteSetBstMidEndSepPunct{\mcitedefaultmidpunct}
{\mcitedefaultendpunct}{\mcitedefaultseppunct}\relax
\EndOfBibitem
\bibitem[Zheng and Zhang(2021)Zheng, and Zhang]{zheng_maggene_2021}
Zheng,~F.; Zhang,~P. {{MagGene}}: {{A}} Genetic Evolution Program for Magnetic Structure Prediction. \emph{Computer Physics Communications} \textbf{2021}, \emph{259}, 107659\relax
\mciteBstWouldAddEndPuncttrue
\mciteSetBstMidEndSepPunct{\mcitedefaultmidpunct}
{\mcitedefaultendpunct}{\mcitedefaultseppunct}\relax
\EndOfBibitem
\bibitem[Baumsteiger \latin{et~al.}(2025)Baumsteiger, Celiberti, Rinke, Todorović, and Franchini]{baumsteiger_exploring_2025}
Baumsteiger,~J.; Celiberti,~L.; Rinke,~P.; Todorović,~M.; Franchini,~C. Exploring noncollinear magnetic energy landscapes with {Bayesian} optimization. \emph{Digital Discovery} \textbf{2025}, \emph{4}, 1639--1650\relax
\mciteBstWouldAddEndPuncttrue
\mciteSetBstMidEndSepPunct{\mcitedefaultmidpunct}
{\mcitedefaultendpunct}{\mcitedefaultseppunct}\relax
\EndOfBibitem
\bibitem[Campi \latin{et~al.}(2022)Campi, Mounet, Gibertini, Pizzi, and Marzari]{campi_materials_2022}
Campi,~D.; Mounet,~N.; Gibertini,~M.; Pizzi,~G.; Marzari,~N. The {{Materials Cloud 2D}} Database ({{MC2D}}). 2022\relax
\mciteBstWouldAddEndPuncttrue
\mciteSetBstMidEndSepPunct{\mcitedefaultmidpunct}
{\mcitedefaultendpunct}{\mcitedefaultseppunct}\relax
\EndOfBibitem
\bibitem[mc2()]{mc2d}
Materials {{Cloud}} Two-Dimensional Crystals Database Table. https://www.materialscloud.org/discover/mc2d/\relax
\mciteBstWouldAddEndPuncttrue
\mciteSetBstMidEndSepPunct{\mcitedefaultmidpunct}
{\mcitedefaultendpunct}{\mcitedefaultseppunct}\relax
\EndOfBibitem
\bibitem[Pizzi \latin{et~al.}(2016)Pizzi, Cepellotti, Sabatini, Marzari, and Kozinsky]{pizzi_aiida_2016}
Pizzi,~G.; Cepellotti,~A.; Sabatini,~R.; Marzari,~N.; Kozinsky,~B. {{AiiDA}}: Automated Interactive Infrastructure and Database for Computational Science. \emph{Computational Materials Science} \textbf{2016}, \emph{111}, 218--230\relax
\mciteBstWouldAddEndPuncttrue
\mciteSetBstMidEndSepPunct{\mcitedefaultmidpunct}
{\mcitedefaultendpunct}{\mcitedefaultseppunct}\relax
\EndOfBibitem
\bibitem[Huber \latin{et~al.}(2020)Huber, Zoupanos, Uhrin, Talirz, Kahle, H{\"a}uselmann, Gresch, M{\"u}ller, Yakutovich, Andersen, Ramirez, Adorf, Gargiulo, Kumbhar, Passaro, Johnston, Merkys, Cepellotti, Mounet, Marzari, Kozinsky, and Pizzi]{huber_aiida_2020}
Huber,~S.~P.; Zoupanos,~S.; Uhrin,~M.; Talirz,~L.; Kahle,~L.; H{\"a}uselmann,~R.; Gresch,~D.; M{\"u}ller,~T.; Yakutovich,~A.~V.; Andersen,~C.~W.; Ramirez,~F.~F.; Adorf,~C.~S.; Gargiulo,~F.; Kumbhar,~S.; Passaro,~E.; Johnston,~C.; Merkys,~A.; Cepellotti,~A.; Mounet,~N.; Marzari,~N. \latin{et~al.}  {{AiiDA}} 1.0, a Scalable Computational Infrastructure for Automated Reproducible Workflows and Data Provenance. \emph{Scientific Data} \textbf{2020}, \emph{7}, 300\relax
\mciteBstWouldAddEndPuncttrue
\mciteSetBstMidEndSepPunct{\mcitedefaultmidpunct}
{\mcitedefaultendpunct}{\mcitedefaultseppunct}\relax
\EndOfBibitem
\bibitem[Cohen \latin{et~al.}(2008)Cohen, {Mori-S{\'a}nchez}, and Yang]{cohen_insights_2008}
Cohen,~A.~J.; {Mori-S{\'a}nchez},~P.; Yang,~W. Insights into {{Current Limitations}} of {{Density Functional Theory}}. \emph{Science} \textbf{2008}, \emph{321}, 792--794\relax
\mciteBstWouldAddEndPuncttrue
\mciteSetBstMidEndSepPunct{\mcitedefaultmidpunct}
{\mcitedefaultendpunct}{\mcitedefaultseppunct}\relax
\EndOfBibitem
\bibitem[Bao \latin{et~al.}(2018)Bao, Gagliardi, and Truhlar]{bao_self-interaction_2018}
Bao,~J.~L.; Gagliardi,~L.; Truhlar,~D.~G. Self-{{Interaction Error}} in {{Density Functional Theory}}: {{An Appraisal}}. \emph{The Journal of Physical Chemistry Letters} \textbf{2018}, \emph{9}, 2353--2358\relax
\mciteBstWouldAddEndPuncttrue
\mciteSetBstMidEndSepPunct{\mcitedefaultmidpunct}
{\mcitedefaultendpunct}{\mcitedefaultseppunct}\relax
\EndOfBibitem
\bibitem[Himmetoglu \latin{et~al.}(2014)Himmetoglu, Floris, De~Gironcoli, and Cococcioni]{himmetoglu_hubbard-corrected_2014}
Himmetoglu,~B.; Floris,~A.; De~Gironcoli,~S.; Cococcioni,~M. Hubbard-Corrected {{DFT}} Energy Functionals: {{The LDA}}+{{U}} Description of Correlated Systems. \emph{International Journal of Quantum Chemistry} \textbf{2014}, \emph{114}, 14--49\relax
\mciteBstWouldAddEndPuncttrue
\mciteSetBstMidEndSepPunct{\mcitedefaultmidpunct}
{\mcitedefaultendpunct}{\mcitedefaultseppunct}\relax
\EndOfBibitem
\bibitem[Mosquera and Wasserman(2014)Mosquera, and Wasserman]{mosquera_derivative_2014}
Mosquera,~M.~A.; Wasserman,~A. Derivative Discontinuities in Density Functional Theory. \emph{Molecular Physics} \textbf{2014}, \emph{112}, 2997--3013\relax
\mciteBstWouldAddEndPuncttrue
\mciteSetBstMidEndSepPunct{\mcitedefaultmidpunct}
{\mcitedefaultendpunct}{\mcitedefaultseppunct}\relax
\EndOfBibitem
\bibitem[Anisimov \latin{et~al.}(1993)Anisimov, Solovyev, Korotin, Czy{\.z}yk, and Sawatzky]{anisimov_density-functional_1993}
Anisimov,~V.~I.; Solovyev,~I.~V.; Korotin,~M.~A.; Czy{\.z}yk,~M.~T.; Sawatzky,~G.~A. Density-Functional Theory and {{NiO}} Photoemission Spectra. \emph{Physical Review B} \textbf{1993}, \emph{48}, 16929--16934\relax
\mciteBstWouldAddEndPuncttrue
\mciteSetBstMidEndSepPunct{\mcitedefaultmidpunct}
{\mcitedefaultendpunct}{\mcitedefaultseppunct}\relax
\EndOfBibitem
\bibitem[Solovyev \latin{et~al.}(1994)Solovyev, Dederichs, and Anisimov]{solovyev_corrected_1994}
Solovyev,~I.~V.; Dederichs,~P.~H.; Anisimov,~V.~I. Corrected Atomic Limit in the Local-Density Approximation and the Electronic Structure of {\emph{d}} Impurities in {{Rb}}. \emph{Physical Review B} \textbf{1994}, \emph{50}, 16861--16871\relax
\mciteBstWouldAddEndPuncttrue
\mciteSetBstMidEndSepPunct{\mcitedefaultmidpunct}
{\mcitedefaultendpunct}{\mcitedefaultseppunct}\relax
\EndOfBibitem
\bibitem[Liechtenstein \latin{et~al.}(1995)Liechtenstein, Anisimov, and Zaanen]{liechtenstein_density-functional_1995}
Liechtenstein,~A.~I.; Anisimov,~V.~I.; Zaanen,~J. Density-Functional Theory and Strong Interactions: {{Orbital}} Ordering in {{Mott-Hubbard}} Insulators. \emph{Physical Review B} \textbf{1995}, \emph{52}, R5467--R5470\relax
\mciteBstWouldAddEndPuncttrue
\mciteSetBstMidEndSepPunct{\mcitedefaultmidpunct}
{\mcitedefaultendpunct}{\mcitedefaultseppunct}\relax
\EndOfBibitem
\bibitem[Perdew \latin{et~al.}(1982)Perdew, Parr, Levy, and Balduz]{perdew_density-functional_1982}
Perdew,~J.~P.; Parr,~R.~G.; Levy,~M.; Balduz,~J.~L. Density-{{Functional Theory}} for {{Fractional Particle Number}}: {{Derivative Discontinuities}} of the {{Energy}}. \emph{Physical Review Letters} \textbf{1982}, \emph{49}, 1691--1694\relax
\mciteBstWouldAddEndPuncttrue
\mciteSetBstMidEndSepPunct{\mcitedefaultmidpunct}
{\mcitedefaultendpunct}{\mcitedefaultseppunct}\relax
\EndOfBibitem
\bibitem[Liang \latin{et~al.}(2021)Liang, Du, Wang, Liu, Wu, and Zhang]{liang_tunable_2021}
Liang,~L.; Du,~S.; Wang,~L.; Liu,~Z.; Wu,~J.; Zhang,~S. Tunable {{Magnetic}} and {{Electronic Properties}} of the {{2D CoO}} {\textsubscript{2}} {{Layer}}. \emph{The Journal of Physical Chemistry C} \textbf{2021}, \emph{125}, 873--877\relax
\mciteBstWouldAddEndPuncttrue
\mciteSetBstMidEndSepPunct{\mcitedefaultmidpunct}
{\mcitedefaultendpunct}{\mcitedefaultseppunct}\relax
\EndOfBibitem
\bibitem[Smolyanyuk \latin{et~al.}(2024)Smolyanyuk, {\v S}mejkal, and Mazin]{smolyanyuk_tool_2024}
Smolyanyuk,~A.; {\v S}mejkal,~L.; Mazin,~I.~I. A Tool to Check Whether a Symmetry-Compensated Collinear Magnetic Material Is Antiferro- or Altermagnetic. \emph{SciPost Physics Codebases} \textbf{2024}, 30\relax
\mciteBstWouldAddEndPuncttrue
\mciteSetBstMidEndSepPunct{\mcitedefaultmidpunct}
{\mcitedefaultendpunct}{\mcitedefaultseppunct}\relax
\EndOfBibitem
\bibitem[S{\o}dequist and Olsen(2024)S{\o}dequist, and Olsen]{sodequist_two-dimensional_2024}
S{\o}dequist,~J.; Olsen,~T. Two-Dimensional Altermagnets from High Throughput Computational Screening: {{Symmetry}} Requirements, Chiral Magnons, and Spin-Orbit Effects. \emph{Applied Physics Letters} \textbf{2024}, \emph{124}, 182409\relax
\mciteBstWouldAddEndPuncttrue
\mciteSetBstMidEndSepPunct{\mcitedefaultmidpunct}
{\mcitedefaultendpunct}{\mcitedefaultseppunct}\relax
\EndOfBibitem
\bibitem[Zeng and Zhao(2024)Zeng, and Zhao]{zeng_description_2024}
Zeng,~S.; Zhao,~Y.-J. Description of Two-Dimensional Altermagnetism: Categorization Using Spin Group Theory. \emph{arXiv preprint arXiv:2405.03557} \textbf{2024}, \emph{2405.03557}, —\relax
\mciteBstWouldAddEndPuncttrue
\mciteSetBstMidEndSepPunct{\mcitedefaultmidpunct}
{\mcitedefaultendpunct}{\mcitedefaultseppunct}\relax
\EndOfBibitem
\bibitem[Chen \latin{et~al.}(2019)Chen, Sun, Wang, Gu, Xu, Wu, and Gao]{chen_direct_2019}
Chen,~W.; Sun,~Z.; Wang,~Z.; Gu,~L.; Xu,~X.; Wu,~S.; Gao,~C. Direct Observation of van Der {{Waals}} Stacking--Dependent Interlayer Magnetism. \emph{Science} \textbf{2019}, \emph{366}, 983--987\relax
\mciteBstWouldAddEndPuncttrue
\mciteSetBstMidEndSepPunct{\mcitedefaultmidpunct}
{\mcitedefaultendpunct}{\mcitedefaultseppunct}\relax
\EndOfBibitem
\bibitem[Lee \latin{et~al.}(2021)Lee, Dismukes, Telford, Wiscons, Wang, Xu, Nuckolls, Dean, Roy, and Zhu]{lee_magnetic_2021}
Lee,~K.; Dismukes,~A.~H.; Telford,~E.~J.; Wiscons,~R.~A.; Wang,~J.; Xu,~X.; Nuckolls,~C.; Dean,~C.~R.; Roy,~X.; Zhu,~X. Magnetic {{Order}} and {{Symmetry}} in the {{2D Semiconductor CrSBr}}. \emph{Nano Letters} \textbf{2021}, \emph{21}, 3511--3517\relax
\mciteBstWouldAddEndPuncttrue
\mciteSetBstMidEndSepPunct{\mcitedefaultmidpunct}
{\mcitedefaultendpunct}{\mcitedefaultseppunct}\relax
\EndOfBibitem
\bibitem[Son \latin{et~al.}(2021)Son, Son, Park, Kim, Tao, Oh, Lee, Lee, Kim, Zhang, Cho, Kamiyama, Lee, Mak, Shan, Kim, Park, and Lee]{son_air-stable_2021}
Son,~J.; Son,~S.; Park,~P.; Kim,~M.; Tao,~Z.; Oh,~J.; Lee,~T.; Lee,~S.; Kim,~J.; Zhang,~K.; Cho,~K.; Kamiyama,~T.; Lee,~J.~H.; Mak,~K.~F.; Shan,~J.; Kim,~M.; Park,~J.-G.; Lee,~J. Air-{{Stable}} and {{Layer-Dependent Ferromagnetism}} in {{Atomically Thin}} van Der {{Waals CrPS}}{\textsubscript{4}}. \emph{ACS Nano} \textbf{2021}, \emph{15}, 16904--16912\relax
\mciteBstWouldAddEndPuncttrue
\mciteSetBstMidEndSepPunct{\mcitedefaultmidpunct}
{\mcitedefaultendpunct}{\mcitedefaultseppunct}\relax
\EndOfBibitem
\bibitem[Kim \latin{et~al.}(2019)Kim, Lim, Lee, Lee, Kim, Park, Jeon, Park, Park, and Cheong]{kim_suppression_2019}
Kim,~K.; Lim,~S.~Y.; Lee,~J.-U.; Lee,~S.; Kim,~T.~Y.; Park,~K.; Jeon,~G.~S.; Park,~C.-H.; Park,~J.-G.; Cheong,~H. Suppression of Magnetic Ordering in {{XXZ-type}} Antiferromagnetic Monolayer {{NiPS}}{\textsubscript{3}}. \emph{Nature Communications} \textbf{2019}, \emph{10}, 345\relax
\mciteBstWouldAddEndPuncttrue
\mciteSetBstMidEndSepPunct{\mcitedefaultmidpunct}
{\mcitedefaultendpunct}{\mcitedefaultseppunct}\relax
\EndOfBibitem
\bibitem[Lee \latin{et~al.}(2016)Lee, Lee, Ryoo, Kang, Kim, Kim, Park, Park, and Cheong]{lee_ising-type_2016}
Lee,~J.-U.; Lee,~S.; Ryoo,~J.~H.; Kang,~S.; Kim,~T.~Y.; Kim,~P.; Park,~C.-H.; Park,~J.-G.; Cheong,~H. Ising-{{Type Magnetic Ordering}} in {{Atomically Thin FePS}}{\textsubscript{3}}. \emph{Nano Letters} \textbf{2016}, \emph{16}, 7433--7438\relax
\mciteBstWouldAddEndPuncttrue
\mciteSetBstMidEndSepPunct{\mcitedefaultmidpunct}
{\mcitedefaultendpunct}{\mcitedefaultseppunct}\relax
\EndOfBibitem
\bibitem[Olsen(2024)]{olsen_antiferromagnetism_2024}
Olsen,~T. Antiferromagnetism in Two-Dimensional Materials: Progress and Computational Challenges. \emph{2D Materials} \textbf{2024}, \emph{11}, 033005\relax
\mciteBstWouldAddEndPuncttrue
\mciteSetBstMidEndSepPunct{\mcitedefaultmidpunct}
{\mcitedefaultendpunct}{\mcitedefaultseppunct}\relax
\EndOfBibitem
\bibitem[Kim \latin{et~al.}(2019)Kim, Kumaravadivel, Birkbeck, Kuang, Xu, Hopkinson, Knolle, McClarty, Berdyugin, Ben~Shalom, Gorbachev, Haigh, Liu, Edgar, Novoselov, Grigorieva, and Geim]{kim_micromagnetometry_2019}
Kim,~M.; Kumaravadivel,~P.; Birkbeck,~J.; Kuang,~W.; Xu,~S.~G.; Hopkinson,~D.~G.; Knolle,~J.; McClarty,~P.~A.; Berdyugin,~A.~I.; Ben~Shalom,~M.; Gorbachev,~R.~V.; Haigh,~S.~J.; Liu,~S.; Edgar,~J.~H.; Novoselov,~K.~S.; Grigorieva,~I.~V.; Geim,~A.~K. Micromagnetometry of Two-Dimensional Ferromagnets. \emph{Nature Electronics} \textbf{2019}, \emph{2}, 457--463\relax
\mciteBstWouldAddEndPuncttrue
\mciteSetBstMidEndSepPunct{\mcitedefaultmidpunct}
{\mcitedefaultendpunct}{\mcitedefaultseppunct}\relax
\EndOfBibitem
\bibitem[Kim \latin{et~al.}(2019)Kim, Yang, Li, Jiang, Jin, Tao, Nichols, Sfigakis, Zhong, Li, Tian, Cory, Miao, Shan, Mak, Lei, Sun, Zhao, and Tsen]{kim_evolution_2019}
Kim,~H.~H.; Yang,~B.; Li,~S.; Jiang,~S.; Jin,~C.; Tao,~Z.; Nichols,~G.; Sfigakis,~F.; Zhong,~S.; Li,~C.; Tian,~S.; Cory,~D.~G.; Miao,~G.-X.; Shan,~J.; Mak,~K.~F.; Lei,~H.; Sun,~K.; Zhao,~L.; Tsen,~A.~W. Evolution of Interlayer and Intralayer Magnetism in Three Atomically Thin Chromium Trihalides. \emph{Proceedings of the National Academy of Sciences} \textbf{2019}, \emph{116}, 11131--11136\relax
\mciteBstWouldAddEndPuncttrue
\mciteSetBstMidEndSepPunct{\mcitedefaultmidpunct}
{\mcitedefaultendpunct}{\mcitedefaultseppunct}\relax
\EndOfBibitem
\bibitem[Zhang \latin{et~al.}(2019)Zhang, Shang, Jiang, Rasmita, Gao, and Yu]{zhang_direct_2019}
Zhang,~Z.; Shang,~J.; Jiang,~C.; Rasmita,~A.; Gao,~W.; Yu,~T. Direct {{Photoluminescence Probing}} of {{Ferromagnetism}} in {{Monolayer Two-Dimensional CrBr}}{\textsubscript{3}}. \emph{Nano Letters} \textbf{2019}, \emph{19}, 3138--3142\relax
\mciteBstWouldAddEndPuncttrue
\mciteSetBstMidEndSepPunct{\mcitedefaultmidpunct}
{\mcitedefaultendpunct}{\mcitedefaultseppunct}\relax
\EndOfBibitem
\bibitem[Torelli \latin{et~al.}(2020)Torelli, Moustafa, Jacobsen, and Olsen]{torelli_high-throughput_2020}
Torelli,~D.; Moustafa,~H.; Jacobsen,~K.~W.; Olsen,~T. High-Throughput Computational Screening for Two-Dimensional Magnetic Materials Based on Experimental Databases of Three-Dimensional Compounds. \emph{npj Computational Materials} \textbf{2020}, \emph{6}, 158\relax
\mciteBstWouldAddEndPuncttrue
\mciteSetBstMidEndSepPunct{\mcitedefaultmidpunct}
{\mcitedefaultendpunct}{\mcitedefaultseppunct}\relax
\EndOfBibitem
\bibitem[Xiang \latin{et~al.}(2011)Xiang, Kan, Wei, Whangbo, and Gong]{xiang_predicting_2011}
Xiang,~H.~J.; Kan,~E.~J.; Wei,~S.-H.; Whangbo,~M.-H.; Gong,~X.~G. Predicting the Spin-Lattice Order of Frustrated Systems from First Principles. \emph{Physical Review B} \textbf{2011}, \emph{84}, 224429\relax
\mciteBstWouldAddEndPuncttrue
\mciteSetBstMidEndSepPunct{\mcitedefaultmidpunct}
{\mcitedefaultendpunct}{\mcitedefaultseppunct}\relax
\EndOfBibitem
\bibitem[Jenkins \latin{et~al.}(2022)Jenkins, R{\'o}zsa, Atxitia, Evans, Novoselov, and Santos]{jenkins_breaking_2022}
Jenkins,~S.; R{\'o}zsa,~L.; Atxitia,~U.; Evans,~R. F.~L.; Novoselov,~K.~S.; Santos,~E. J.~G. Breaking through the {{Mermin-Wagner}} Limit in {{2D}} van Der {{Waals}} Magnets. \emph{Nature Communications} \textbf{2022}, \emph{13}, 6917\relax
\mciteBstWouldAddEndPuncttrue
\mciteSetBstMidEndSepPunct{\mcitedefaultmidpunct}
{\mcitedefaultendpunct}{\mcitedefaultseppunct}\relax
\EndOfBibitem
\bibitem[Garanin(1996)]{garanin_self-consistent_1996}
Garanin,~D.~A. Self-Consistent {{Gaussian}} Approximation for Classical Spin Systems: {{Thermodynamics}}. \emph{Physical Review B} \textbf{1996}, \emph{53}, 11593--11605\relax
\mciteBstWouldAddEndPuncttrue
\mciteSetBstMidEndSepPunct{\mcitedefaultmidpunct}
{\mcitedefaultendpunct}{\mcitedefaultseppunct}\relax
\EndOfBibitem
\bibitem[Pizzochero \latin{et~al.}(2020)Pizzochero, Yadav, and Yazyev]{pizzochero_magnetic_2020}
Pizzochero,~M.; Yadav,~R.; Yazyev,~O.~V. Magnetic Exchange Interactions in Monolayer {{CrI}} {\textsubscript{3}} from Many-Body Wavefunction Calculations. \emph{2D Materials} \textbf{2020}, \emph{7}, 035005\relax
\mciteBstWouldAddEndPuncttrue
\mciteSetBstMidEndSepPunct{\mcitedefaultmidpunct}
{\mcitedefaultendpunct}{\mcitedefaultseppunct}\relax
\EndOfBibitem
\bibitem[Li and Yang(2016)Li, and Yang]{li_first-principles_2016}
Li,~X.; Yang,~J. First-Principles Design of Spintronics Materials. \emph{National Science Review} \textbf{2016}, \emph{3}, 365--381\relax
\mciteBstWouldAddEndPuncttrue
\mciteSetBstMidEndSepPunct{\mcitedefaultmidpunct}
{\mcitedefaultendpunct}{\mcitedefaultseppunct}\relax
\EndOfBibitem
\bibitem[De~Groot \latin{et~al.}(1983)De~Groot, Mueller, Engen, and Buschow]{de_groot_new_1983}
De~Groot,~R.~A.; Mueller,~F.~M.; Engen,~P. G.~V.; Buschow,~K. H.~J. New {{Class}} of {{Materials}}: {{Half-Metallic Ferromagnets}}. \emph{Physical Review Letters} \textbf{1983}, \emph{50}, 2024--2027\relax
\mciteBstWouldAddEndPuncttrue
\mciteSetBstMidEndSepPunct{\mcitedefaultmidpunct}
{\mcitedefaultendpunct}{\mcitedefaultseppunct}\relax
\EndOfBibitem
\bibitem[Yao \latin{et~al.}(2021)Yao, Li, and Liu]{yao_fragile_2021}
Yao,~Q.; Li,~J.; Liu,~Q. Fragile Symmetry-Protected Half Metallicity in Two-Dimensional van Der {{Waals}} Magnets: {{A}} Case Study of Monolayer {{Fe Cl}} 2. \emph{Physical Review B} \textbf{2021}, \emph{104}, 035108\relax
\mciteBstWouldAddEndPuncttrue
\mciteSetBstMidEndSepPunct{\mcitedefaultmidpunct}
{\mcitedefaultendpunct}{\mcitedefaultseppunct}\relax
\EndOfBibitem
\bibitem[Ashton \latin{et~al.}(2017)Ashton, Gluhovic, Sinnott, Guo, Stewart, and Hennig]{ashton_two-dimensional_2017}
Ashton,~M.; Gluhovic,~D.; Sinnott,~S.~B.; Guo,~J.; Stewart,~D.~A.; Hennig,~R.~G. Two-{{Dimensional Intrinsic Half-Metals With Large Spin Gaps}}. \emph{Nano Letters} \textbf{2017}, \emph{17}, 5251--5257\relax
\mciteBstWouldAddEndPuncttrue
\mciteSetBstMidEndSepPunct{\mcitedefaultmidpunct}
{\mcitedefaultendpunct}{\mcitedefaultseppunct}\relax
\EndOfBibitem
\bibitem[Torun \latin{et~al.}(2015)Torun, Sahin, Singh, and Peeters]{torun_stable_2015}
Torun,~E.; Sahin,~H.; Singh,~S.~K.; Peeters,~F.~M. Stable Half-Metallic Monolayers of {{FeCl2}}. \emph{Applied Physics Letters} \textbf{2015}, \emph{106}, 192404\relax
\mciteBstWouldAddEndPuncttrue
\mciteSetBstMidEndSepPunct{\mcitedefaultmidpunct}
{\mcitedefaultendpunct}{\mcitedefaultseppunct}\relax
\EndOfBibitem
\bibitem[Hadjadj \latin{et~al.}(2023)Hadjadj, {Gonz{\'a}lez-Orellana}, Lawrence, Bikaljevi{\'c}, {Pe{\~n}a-D{\'i}az}, Gargiani, Aballe, Naumann, Ni{\~n}o, Foerster, {Ruiz-G{\'o}mez}, Thakur, Kumberg, Taylor, Hayes, Torres, Luo, Radu, De~Oteyza, Kuch, Pascual, Rogero, and Ilyn]{hadjadj_epitaxial_2023}
Hadjadj,~S.~E.; {Gonz{\'a}lez-Orellana},~C.; Lawrence,~J.; Bikaljevi{\'c},~D.; {Pe{\~n}a-D{\'i}az},~M.; Gargiani,~P.; Aballe,~L.; Naumann,~J.; Ni{\~n}o,~M.~{\'A}.; Foerster,~M.; {Ruiz-G{\'o}mez},~S.; Thakur,~S.; Kumberg,~I.; Taylor,~J.~M.; Hayes,~J.; Torres,~J.; Luo,~C.; Radu,~F.; De~Oteyza,~D.~G.; Kuch,~W. \latin{et~al.}  Epitaxial {{Monolayers}} of the {{Magnetic 2D Semiconductor FeBr}} {\textsubscript{2}} {{Grown}} on {{Au}}(111). \emph{Chemistry of Materials} \textbf{2023}, \emph{35}, 9847--9856\relax
\mciteBstWouldAddEndPuncttrue
\mciteSetBstMidEndSepPunct{\mcitedefaultmidpunct}
{\mcitedefaultendpunct}{\mcitedefaultseppunct}\relax
\EndOfBibitem
\bibitem[Prayitno(2021)]{prayitno_controlling_2021}
Prayitno,~T.~B. Controlling Phase Transition in Monolayer Metal Diiodides {{XI}} {\textsubscript{2}} ({{X}}: {{Fe}}, {{Co}}, and {{Ni}}) by Carrier Doping. \emph{Journal of Physics: Condensed Matter} \textbf{2021}, \emph{33}, 335803\relax
\mciteBstWouldAddEndPuncttrue
\mciteSetBstMidEndSepPunct{\mcitedefaultmidpunct}
{\mcitedefaultendpunct}{\mcitedefaultseppunct}\relax
\EndOfBibitem
\bibitem[Zhou \latin{et~al.}(2024)Zhou, Jiang, Tao, Ji, Wang, Lai, and Zhong]{zhou_evidence_2024}
Zhou,~X.; Jiang,~T.; Tao,~Y.; Ji,~Y.; Wang,~J.; Lai,~T.; Zhong,~D. Evidence of {{Ferromagnetism}} and {{Ultrafast Dynamics}} of {{Demagnetization}} in an {{Epitaxial FeCl}} {\textsubscript{2}} {{Monolayer}}. \emph{ACS Nano} \textbf{2024}, \emph{18}, 10912--10920\relax
\mciteBstWouldAddEndPuncttrue
\mciteSetBstMidEndSepPunct{\mcitedefaultmidpunct}
{\mcitedefaultendpunct}{\mcitedefaultseppunct}\relax
\EndOfBibitem
\bibitem[Kong \latin{et~al.}(2020)Kong, Li, Liang, Peeters, and Liu]{kong_magnetic_2020}
Kong,~X.; Li,~L.; Liang,~L.; Peeters,~F.~M.; Liu,~X.-J. The Magnetic, Electronic, and Light-Induced Topological Properties in Two-Dimensional Hexagonal {{FeX2}} ({{X}} = {{Cl}}, {{Br}}, {{I}}) Monolayers. \emph{Applied Physics Letters} \textbf{2020}, \emph{116}, 192404\relax
\mciteBstWouldAddEndPuncttrue
\mciteSetBstMidEndSepPunct{\mcitedefaultmidpunct}
{\mcitedefaultendpunct}{\mcitedefaultseppunct}\relax
\EndOfBibitem
\bibitem[Cai \latin{et~al.}(2020)Cai, Yang, and Gao]{cai_fecl2_2020}
Cai,~S.; Yang,~F.; Gao,~C. {{FeCl}}{\textsubscript{2}} Monolayer on {{HOPG}}: Art of Growth and Momentum Filtering Effect. \emph{Nanoscale} \textbf{2020}, \emph{12}, 16041--16045\relax
\mciteBstWouldAddEndPuncttrue
\mciteSetBstMidEndSepPunct{\mcitedefaultmidpunct}
{\mcitedefaultendpunct}{\mcitedefaultseppunct}\relax
\EndOfBibitem
\bibitem[Giannozzi \latin{et~al.}(2009)Giannozzi, Baroni, Bonini, Calandra, Car, Cavazzoni, Ceresoli, Chiarotti, Cococcioni, Dabo, Dal~Corso, {de Gironcoli}, Fabris, Fratesi, Gebauer, Gerstmann, Gougoussis, Kokalj, Lazzeri, {Martin-Samos}, Marzari, Mauri, Mazzarello, Paolini, Pasquarello, Paulatto, Sbraccia, Scandolo, Sclauzero, Seitsonen, Smogunov, Umari, and Wentzcovitch]{giannozzi_quantum_2009}
Giannozzi,~P.; Baroni,~S.; Bonini,~N.; Calandra,~M.; Car,~R.; Cavazzoni,~C.; Ceresoli,~D.; Chiarotti,~G.~L.; Cococcioni,~M.; Dabo,~I.; Dal~Corso,~A.; {de Gironcoli},~S.; Fabris,~S.; Fratesi,~G.; Gebauer,~R.; Gerstmann,~U.; Gougoussis,~C.; Kokalj,~A.; Lazzeri,~M.; {Martin-Samos},~L. \latin{et~al.}  {{QUANTUM ESPRESSO}}: A Modular and Open-Source Software Project for Quantum Simulations of Materials. \emph{Journal of Physics: Condensed Matter} \textbf{2009}, \emph{21}, 395502\relax
\mciteBstWouldAddEndPuncttrue
\mciteSetBstMidEndSepPunct{\mcitedefaultmidpunct}
{\mcitedefaultendpunct}{\mcitedefaultseppunct}\relax
\EndOfBibitem
\bibitem[Giannozzi \latin{et~al.}(2017)Giannozzi, Andreussi, Brumme, Bunau, Buongiorno~Nardelli, Calandra, Car, Cavazzoni, Ceresoli, Cococcioni, Colonna, Carnimeo, Dal~Corso, {de Gironcoli}, Delugas, DiStasio, Ferretti, Floris, Fratesi, Fugallo, Gebauer, Gerstmann, Giustino, Gorni, Jia, Kawamura, Ko, Kokalj, K{\"u}{\c c}{\"u}kbenli, Lazzeri, Marsili, Marzari, Mauri, Nguyen, Nguyen, {Otero-de-la-Roza}, Paulatto, Ponc{\'e}, Rocca, Sabatini, Santra, Schlipf, Seitsonen, Smogunov, Timrov, Thonhauser, Umari, Vast, Wu, and Baroni]{giannozzi_advanced_2017}
Giannozzi,~P.; Andreussi,~O.; Brumme,~T.; Bunau,~O.; Buongiorno~Nardelli,~M.; Calandra,~M.; Car,~R.; Cavazzoni,~C.; Ceresoli,~D.; Cococcioni,~M.; Colonna,~N.; Carnimeo,~I.; Dal~Corso,~A.; {de Gironcoli},~S.; Delugas,~P.; DiStasio,~R.~A.; Ferretti,~A.; Floris,~A.; Fratesi,~G.; Fugallo,~G. \latin{et~al.}  Advanced Capabilities for Materials Modelling with {{Quantum ESPRESSO}}. \emph{Journal of Physics: Condensed Matter} \textbf{2017}, \emph{29}, 465901\relax
\mciteBstWouldAddEndPuncttrue
\mciteSetBstMidEndSepPunct{\mcitedefaultmidpunct}
{\mcitedefaultendpunct}{\mcitedefaultseppunct}\relax
\EndOfBibitem
\bibitem[Giannozzi \latin{et~al.}(2020)Giannozzi, Baseggio, Bonf{\`a}, Brunato, Car, Carnimeo, Cavazzoni, {de Gironcoli}, Delugas, Ferrari~Ruffino, Ferretti, Marzari, Timrov, Urru, and Baroni]{giannozzi_q_2020}
Giannozzi,~P.; Baseggio,~O.; Bonf{\`a},~P.; Brunato,~D.; Car,~R.; Carnimeo,~I.; Cavazzoni,~C.; {de Gironcoli},~S.; Delugas,~P.; Ferrari~Ruffino,~F.; Ferretti,~A.; Marzari,~N.; Timrov,~I.; Urru,~A.; Baroni,~S. Q {\textsc{Uantum}} {{ESPRESSO}} toward the Exascale. \emph{The Journal of Chemical Physics} \textbf{2020}, \emph{152}, 154105\relax
\mciteBstWouldAddEndPuncttrue
\mciteSetBstMidEndSepPunct{\mcitedefaultmidpunct}
{\mcitedefaultendpunct}{\mcitedefaultseppunct}\relax
\EndOfBibitem
\bibitem[Prandini \latin{et~al.}(2018)Prandini, Marrazzo, Castelli, Mounet, and Marzari]{prandini_precision_2018}
Prandini,~G.; Marrazzo,~A.; Castelli,~I.~E.; Mounet,~N.; Marzari,~N. Precision and Efficiency in Solid-State Pseudopotential Calculations. \emph{npj Computational Materials} \textbf{2018}, \emph{4}, 72\relax
\mciteBstWouldAddEndPuncttrue
\mciteSetBstMidEndSepPunct{\mcitedefaultmidpunct}
{\mcitedefaultendpunct}{\mcitedefaultseppunct}\relax
\EndOfBibitem
\bibitem[Lejaeghere \latin{et~al.}(2016)Lejaeghere, Bihlmayer, Bj{\"o}rkman, Blaha, Bl{\"u}gel, Blum, Caliste, Castelli, Clark, Dal~Corso, {de Gironcoli}, Deutsch, Dewhurst, Di~Marco, Draxl, Du{\l}ak, Eriksson, {Flores-Livas}, Garrity, Genovese, Giannozzi, Giantomassi, Goedecker, Gonze, Gr{\aa}n{\"a}s, Gross, Gulans, Gygi, Hamann, Hasnip, Holzwarth, Iu{\c s}an, Jochym, Jollet, Jones, Kresse, Koepernik, K{\"u}{\c c}{\"u}kbenli, Kvashnin, Locht, Lubeck, Marsman, Marzari, Nitzsche, Nordstr{\"o}m, Ozaki, Paulatto, Pickard, Poelmans, Probert, Refson, Richter, Rignanese, Saha, Scheffler, Schlipf, Schwarz, Sharma, Tavazza, Thunstr{\"o}m, Tkatchenko, Torrent, Vanderbilt, {van Setten}, Van~Speybroeck, Wills, Yates, Zhang, and Cottenier]{lejaeghere_reproducibility_2016}
Lejaeghere,~K.; Bihlmayer,~G.; Bj{\"o}rkman,~T.; Blaha,~P.; Bl{\"u}gel,~S.; Blum,~V.; Caliste,~D.; Castelli,~I.~E.; Clark,~S.~J.; Dal~Corso,~A.; {de Gironcoli},~S.; Deutsch,~T.; Dewhurst,~J.~K.; Di~Marco,~I.; Draxl,~C.; Du{\l}ak,~M.; Eriksson,~O.; {Flores-Livas},~J.~A.; Garrity,~K.~F.; Genovese,~L. \latin{et~al.}  Reproducibility in Density Functional Theory Calculations of Solids. \emph{Science} \textbf{2016}, \emph{351}, aad3000\relax
\mciteBstWouldAddEndPuncttrue
\mciteSetBstMidEndSepPunct{\mcitedefaultmidpunct}
{\mcitedefaultendpunct}{\mcitedefaultseppunct}\relax
\EndOfBibitem
\bibitem[Vanderbilt(1990)]{vanderbilt_soft_1990}
Vanderbilt,~D. Soft Self-Consistent Pseudopotentials in a Generalized Eigenvalue Formalism. \emph{Physical Review B} \textbf{1990}, \emph{41}, 7892--7895\relax
\mciteBstWouldAddEndPuncttrue
\mciteSetBstMidEndSepPunct{\mcitedefaultmidpunct}
{\mcitedefaultendpunct}{\mcitedefaultseppunct}\relax
\EndOfBibitem
\bibitem[Dal~Corso(2014)]{dal_corso_pseudopotentials_2014}
Dal~Corso,~A. Pseudopotentials Periodic Table: {{From H}} to {{Pu}}. \emph{Computational Materials Science} \textbf{2014}, \emph{95}, 337--350\relax
\mciteBstWouldAddEndPuncttrue
\mciteSetBstMidEndSepPunct{\mcitedefaultmidpunct}
{\mcitedefaultendpunct}{\mcitedefaultseppunct}\relax
\EndOfBibitem
\bibitem[Garrity \latin{et~al.}(2014)Garrity, Bennett, Rabe, and Vanderbilt]{garrity_pseudopotentials_2014}
Garrity,~K.~F.; Bennett,~J.~W.; Rabe,~K.~M.; Vanderbilt,~D. Pseudopotentials for High-Throughput {{DFT}} Calculations. \emph{Computational Materials Science} \textbf{2014}, \emph{81}, 446--452\relax
\mciteBstWouldAddEndPuncttrue
\mciteSetBstMidEndSepPunct{\mcitedefaultmidpunct}
{\mcitedefaultendpunct}{\mcitedefaultseppunct}\relax
\EndOfBibitem
\bibitem[mat()]{materialscloud_arxiv}
Exploring the magnetic landscape of easily-exfoliable two-dimensional materials. https://archive.materialscloud.org/records/4x37g-nre30\relax
\mciteBstWouldAddEndPuncttrue
\mciteSetBstMidEndSepPunct{\mcitedefaultmidpunct}
{\mcitedefaultendpunct}{\mcitedefaultseppunct}\relax
\EndOfBibitem
\bibitem[noa()]{noauthor_inorganic_nodate}
Inorganic {{Crystal Structure Database}} -- {{ICSD}} {\textbar} {{FIZ Karlsruhe}}. https://www.fiz-karlsruhe.de/en/produkte-und-dienstleistungen/inorganic-crystal-structure-database-icsd\relax
\mciteBstWouldAddEndPuncttrue
\mciteSetBstMidEndSepPunct{\mcitedefaultmidpunct}
{\mcitedefaultendpunct}{\mcitedefaultseppunct}\relax
\EndOfBibitem
\bibitem[Bergerhoff \latin{et~al.}(1983)Bergerhoff, Hundt, Sievers, and Brown]{bergerhoff_inorganic_1983}
Bergerhoff,~G.; Hundt,~R.; Sievers,~R.; Brown,~I.~D. The Inorganic Crystal Structure Data Base. \emph{Journal of Chemical Information and Computer Sciences} \textbf{1983}, \emph{23}, 66--69\relax
\mciteBstWouldAddEndPuncttrue
\mciteSetBstMidEndSepPunct{\mcitedefaultmidpunct}
{\mcitedefaultendpunct}{\mcitedefaultseppunct}\relax
\EndOfBibitem
\bibitem[Gra{\v z}ulis \latin{et~al.}(2012)Gra{\v z}ulis, Da{\v s}kevi{\v c}, Merkys, Chateigner, Lutterotti, Quir{\'o}s, Serebryanaya, Moeck, Downs, and Le~Bail]{grazulis_crystallography_2012}
Gra{\v z}ulis,~S.; Da{\v s}kevi{\v c},~A.; Merkys,~A.; Chateigner,~D.; Lutterotti,~L.; Quir{\'o}s,~M.; Serebryanaya,~N.~R.; Moeck,~P.; Downs,~R.~T.; Le~Bail,~A. Crystallography {{Open Database}} ({{COD}}): An Open-Access Collection of Crystal Structures and Platform for World-Wide Collaboration. \emph{Nucleic Acids Research} \textbf{2012}, \emph{40}, D420--D427\relax
\mciteBstWouldAddEndPuncttrue
\mciteSetBstMidEndSepPunct{\mcitedefaultmidpunct}
{\mcitedefaultendpunct}{\mcitedefaultseppunct}\relax
\EndOfBibitem
\bibitem[noa()]{noauthor_materials_nodate}
Materials {{Platform}} for {{Data Science}}. https://mpds.io/\#start\relax
\mciteBstWouldAddEndPuncttrue
\mciteSetBstMidEndSepPunct{\mcitedefaultmidpunct}
{\mcitedefaultendpunct}{\mcitedefaultseppunct}\relax
\EndOfBibitem
\bibitem[Evans \latin{et~al.}(2014)Evans, Fan, Chureemart, Ostler, Ellis, and Chantrell]{evans_atomistic_2014}
Evans,~R. F.~L.; Fan,~W.~J.; Chureemart,~P.; Ostler,~T.~A.; Ellis,~M. O.~A.; Chantrell,~R.~W. Atomistic Spin Model Simulations of Magnetic Nanomaterials. \emph{Journal of Physics: Condensed Matter} \textbf{2014}, \emph{26}, 103202\relax
\mciteBstWouldAddEndPuncttrue
\mciteSetBstMidEndSepPunct{\mcitedefaultmidpunct}
{\mcitedefaultendpunct}{\mcitedefaultseppunct}\relax
\EndOfBibitem
\bibitem[noa()]{noauthor_vampire_nodate}
Vampire. https://vampire.york.ac.uk/\relax
\mciteBstWouldAddEndPuncttrue
\mciteSetBstMidEndSepPunct{\mcitedefaultmidpunct}
{\mcitedefaultendpunct}{\mcitedefaultseppunct}\relax
\EndOfBibitem
\end{mcitethebibliography}

\end{document}